\documentclass[aps,prc,preprint,groupedaddress,amsmath,amssymb,tightenlines]{revtex4-1}

\pdfoutput=1
\usepackage{graphicx}
\usepackage{textcomp}
\usepackage{hyperref}

\begin{document}

\title{Neutron-neutron quasifree scattering in neutron-deuteron breakup  at 10 MeV}
\author{R.~C.~Malone}
\email{rm216@duke.edu}

\author{A.~S.~Crowell}

\author{L.~C.~Cumberbatch}

\author{B.~A.~Fallin}

\author{F.~Q.~L.~Friesen}

\author{C.~R.~Howell}

\author{C.~R.~Malone}

\author{D.~R.~Ticehurst}

\author{W.~Tornow}
\affiliation{Department of Physics, Duke University and Triangle
  Universities Nuclear Laboratory, Durham, NC 27708, USA }

\author{D.~M.~Markoff}

\author{B.~J.~Crowe}
\affiliation{Department of Mathematics and Physics, North Carolina
  Central University Durham, North Carolina 27707 and
  Triangle Universities Nuclear Laboratory,
  Durham, North Carolina 27708, USA }

\author{H.~Wita{\l}a}
\affiliation{M. Smoluchowski Institute of
  Physics, Jagiellonian University, Krak\`{o}w, Poland}

\date{\today}

\begin{abstract}
New measurements of the neutron-neutron quasifree scattering cross
section in neutron-deuteron breakup at an incident neutron energy of
10.0 MeV are reported. The experiment setup was optimized to evaluate
the technique for determining the integrated beam-target luminosity in
neutron-neutron coincidence cross-section measurements in
neutron-deuteron breakup. The measurements were carried out with a
systematic uncertainty of $\pm5.6\%$. Our data are in agreement with
theoretical calculations performed using the CD-Bonn nucleon-nucleon
potential in the Faddeev formalism. The measured integrated cross
section over the quasifree peak is $20.5 \pm 0.5 \, \text{(stat)} \pm
1.1 \, \text{(sys)} \, \text{mb/sr}^{2}$ in comparison with the theory
prediction of 20.1 mb/sr$^{2}$. These results validate our technique
for determining the beam-target luminosity in neutron-deuteron breakup
measurements.
\end{abstract}

\maketitle

\section{Introduction}

The neutron-deuteron (\textit{nd}) system is a robust platform for
testing models of nucleon interactions. Current calculations using
ab-initio methods with state-of-the-art nucleon-nucleon (\textit{NN})
potentials accurately predict most three-nucleon ($3N$) scattering
observables \cite{Glo96}. However, some discrepancies between theory
and data remain, such as for the neutron-neutron quasifree scattering
(\textit{nn} QFS) cross section in \textit{nd} breakup
\cite{Sie02,Rua07,Lub92,Wit11}.

Neutron-neutron QFS in \textit{nd} breakup is the kinematic
configuration in which the proton remains at rest in the laboratory
frame during the scattering process. That is, the proton may be
considered as a spectator to the interaction between the two
neutrons. Ab-initio calculations illustrate that the \textit{nn} QFS
cross section is sensitive to the details of the \textit{nn}
interaction, even at low energies where the de Broglie wavelength of
the incident neutron is comparable in size to the deuteron. This cross
section depends on the \textit{nn} effective range parameter
($r_{nn}$) in the low-momentum expansion of the $s$-wave scattering
amplitude \cite{Wit11}. However, early measurements did not determine
$r_{nn}$ with high enough precision to examine the validity of charge
symmetry in the \textit{NN} interaction
\cite{Sla71,Bov78,Sou79,Gur80,Wit80}.

The situation is significantly changed by recent cross-section
measurements of \textit{nn} QFS in \textit{nd} breakup at incident
neutron energies of 26 and 25 MeV. Rigorous \textit{nd} breakup
calculations underpredict these data by 18\% and 16\%, respectively
\cite{Sie02,Rua07}. A third and earlier experiment measured a similar
discrepancy of 12\% at 10.3 MeV \cite{Lub92}. A detailed analysis of
the 26-MeV \textit{nn} QFS data \cite{Sie02} using rigorous
\textit{nd} breakup calculations demonstrated that $3N$ forces cannot
account for the discrepancy between data and theory
\cite{Wit11}. Also, the analysis showed that theory can be brought
into agreement with data by scaling the magnitude of the ${^1}S_{0}$
\textit{nn} interaction by a factor of 1.08. However, this remedy
suggests substantial charge symmetry breaking in the \textit{NN}
interaction manifested as either: changes to the $^1S_0$ \textit{nn}
scattering length ($a_{nn}$) to the extent of nearly creating a bound
dineutron state, a significant deviation of $r_{nn}$ from the accepted
value of the \textit{NN} effective range parameter, or a combination
of changes to the nominal values of $a_{nn}$ and $r_{nn}$
\cite{Wit11}. Possible explanations for the \textit{nn} QFS
discrepancy include: (1) the \textit{NN} system violates charge
symmetry at a level larger than generally accepted, (2) current $3N$
force models do not properly account for all $3N$ force components
that contribute to the reaction dynamics, and/or (3) the systematic
uncertainties were underestimated in the reported measurements. 

\begin{figure}[htb]
\centering
\includegraphics[width=.9\linewidth]{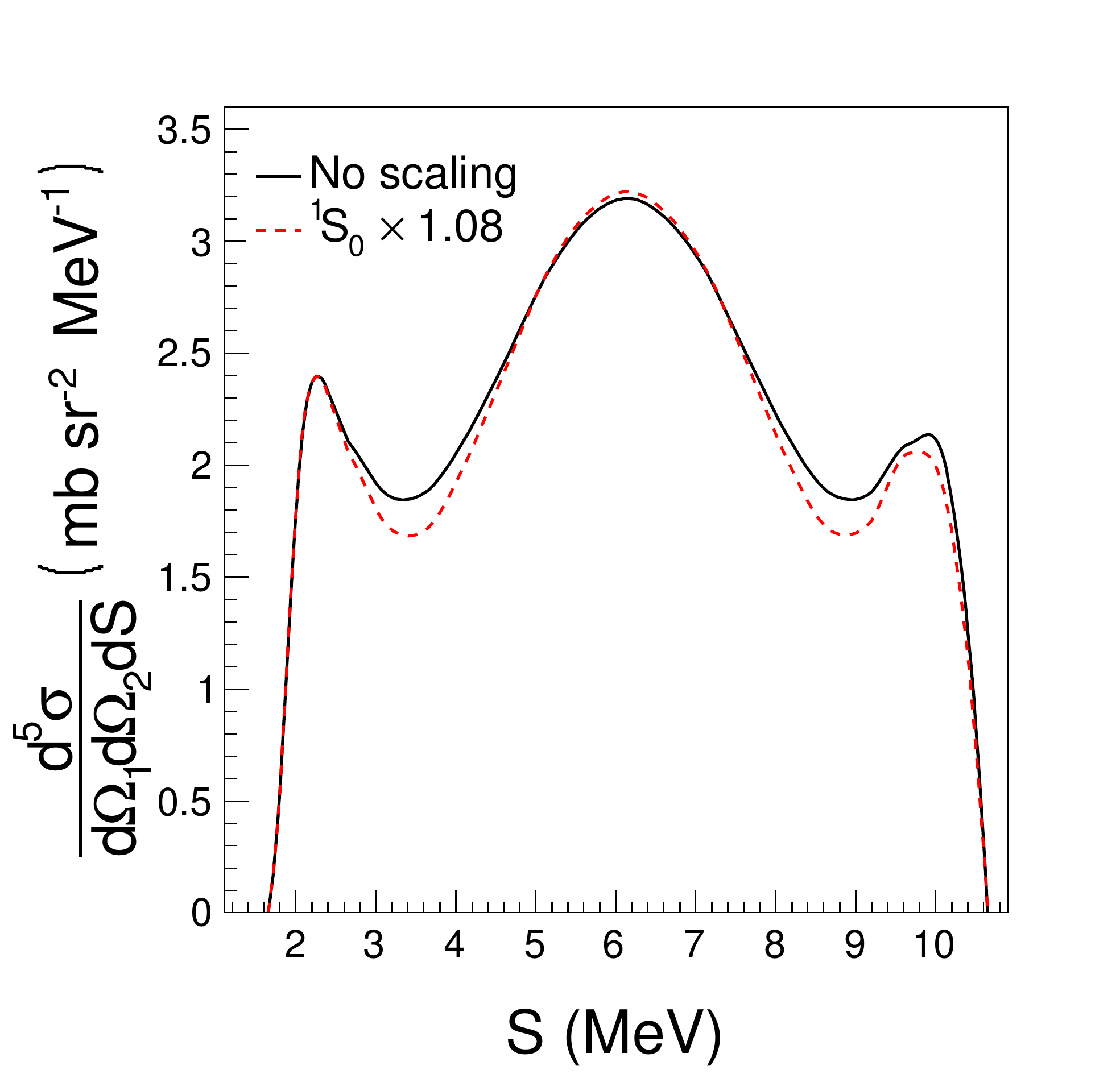}
\caption{\label{fig:nnqfs-10MeV-sensitivity} Plot of the theoretical
  cross section for \textit{nn} QFS in \textit{nd} breakup as a
  function of arc length along the kinematic locus (\textit{S}) for an
  incident neutron energy of $E_{n} = 10.0$ MeV and scattering angles
  of $\theta_{1} = \theta_{2} = 36.7^{\circ}$ and $\Delta \phi =
  180^{\circ}$. Calculations were performed with unscaled interaction
  matrix elements (solid black curve) and with the $^{1}S_{0}$
  \textit{nn} matrix elements scaled by a factor of 1.08 (dashed red
  curve). Both theory calculations have been averaged over the finite
  geometry of the experiment using the Monte-Carlo simulation
  described in this paper. }
\end{figure}

A common feature of the comparisons of theory to data is that
calculations describe the shape of the cross-section distribution
along the kinematic locus well but fail to predict the absolute
magnitude of the data. This type of discrepancy is suggestive that the
systematic uncertainty in the factors used to normalize the
cross-section measurements might be underestimated. That is, an
uncertainty of $\pm18\%$ in the beam-target luminosity would bring
measurements and theory into agreement within one standard
deviation. In this paper, we report new \textit{nn} QFS cross-section
measurements in \textit{nd} breakup. Our experiment method differs
from previous measurements \cite{Sie02,Rua07,Lub92} in that the setup
was optimized to evaluate the technique for measuring the absolute
\textit{nn} QFS cross section in \textit{nd} breakup rather than for
sensitivity studies of the strength of the \textit{nn}
interaction. Another important difference is the method used to
determine the integrated beam-target luminosity. In our experiment,
the beam-target luminosity is determined from in-situ
measurements of the yields for \textit{nd} elastic scattering rather
than from neutron-proton (\textit{np}) scattering. This
technique significantly reduces systematic error in the breakup cross
section in comparison to previous \textit{nn} QFS measurements. Our
measurement was conducted at a neutron beam energy of 10.0 MeV, where
theory predicts that the \textit{nn} QFS cross section measured in the
geometry of our experiment has only modest sensitivity to the
$^{1}S_{0}$ \textit{nn} interaction. That sensitivity is shown in
Fig. \ref{fig:nnqfs-10MeV-sensitivity} where the theoretical cross
section averaged over the finite geometry of our experiment is shown
for calculations with and without scaling the $^{1}S_{0}$ \textit{nn}
interaction by 1.08. The difference in the predicted cross section at
the location of the QFS peak (S = 6 MeV) is only 1\%. Additionally, a
concurrent measurement of the integrated neutron beam flux was made
using \textit{np} scattering to assess the systematic error in
determining the luminosity via \textit{nd} elastic scattering. 

In this paper, we report the results of the 10-MeV measurement. We
describe the setup of the experiment in Sec. \ref{sec:setup}. In
Sec. \ref{sec:analysis} we discuss the details of the data
analysis. Our results are presented in Sec. \ref{sec:results} and
summarized in Sec. \ref{sec:conclusion}.

\section{Details of the Experiment}
\label{sec:setup}

The measurements were conducted at the tandem accelerator facility of
the Triangle Universities Nuclear Laboratory (TUNL) using standard
neutron time-of-flight (TOF) techniques. The neutron beam was produced
via the ${^2}\text{H(}d,n\text{)}{^3}\text{He}$ reaction with a pulsed
deuteron beam (period$\,=400$ ns, FWHM$\,=2$ ns) incident on a
3.16-cm-long gas cell filled with deuterium to a pressure of 5
atm. The resulting neutron beam had a central energy of 10.0 MeV with
a spread of 330 keV (full width) due to energy loss by the deuterons
in the deuterium gas. The deuteron beam current on target was adjusted
to optimize the ratio of the true \textit{nn} coincidence rate to the
accidental coincidence background rate.

\begin{figure}[htb]
\centering
\includegraphics[width=.9\linewidth]{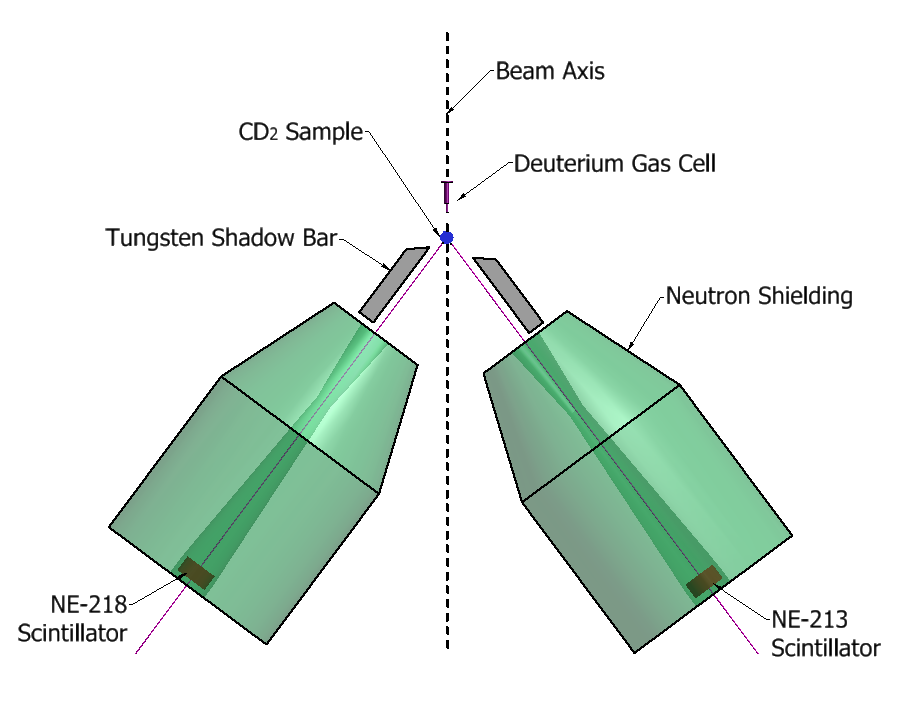}
\caption{\label{fig:ntof-setup}A diagram of the experiment setup
  (distances are to scale). The sample is 12.1 cm from the center of
  the neutron production cell, and the detectors are placed about 265
  cm from the target at 36.7\textdegree{} on either side of the
  beam. More details are given in the text.}
\end{figure}

The experiment setup is shown in Fig. \ref{fig:ntof-setup}. A
cylindrical scattering sample was mounted 12.1 cm from the center of
the gas cell with its axis vertical and centered in the beam at the
location of the pivot point about which the detectors rotate.
Scattered neutrons were detected by two heavily shielded liquid
scintillators positioned on opposite sides of the beam axis at equal
angles of 36.7\textdegree{}. The left and right detectors are 5.08 cm
long cylinders with diameters of 12.7 cm and 8.89 cm filled with
NE-213 and NE-218 liquid scintillator fluid, respectively. Each
detector was housed inside a cylindrical shielding enclosure of
lithium-doped paraffin with a double-truncated conical copper
collimator \cite{Gla74}. Tungsten shadow bars were positioned to
shield the detectors from directly viewing the neutron production
cell. The distance from the center of the sample to the center of each
detector was 264.9 cm for the left detector and 264.3 cm for the right
detector. The neutron beam flux was monitored using two liquid
scintillators not shown in Fig. \ref{fig:ntof-setup}. One monitor
detector (5.08 cm diameter $\times$ 5.08 cm long) was suspended from
the ceiling in a copper shield and collimator and viewed the neutron
production cell at an angle of approximately 60\textdegree{} with
respect to the beam axis. The other detector (3.81 cm diameter
$\times$ 3.81 cm long) was unshielded and positioned approximately
three meters downstream from the neutron production gas cell at an
angle of about 3\textdegree{} relative to the beam axis.

Data were collected over the course of three runs for a total of 577
hours of beam on target. The integrated beam-target luminosity was
determined from the \textit{nd} elastic scattering yields, which were
measured simultaneously with the data for the breakup reaction. The
integrated incident beam flux was also measured using \textit{np}
scattering to check the systematic uncertainty in our determination of
the beam-target luminosity.

All scattering samples used in these measurements were right
cylinders. The mass and dimensions of each sample are given in Table
\ref{tab:sample-dimensions}. The deuterium sample was composed of
98.4\% isotopically enriched deuterated polyethylene, CD$_2$, where
``D'' denotes ``$^2$H''
(Cambridge Isotope Laboratories, Inc., DLM-220-0). The \textit{np}
scattering measurements were performed using the polyethylene (CH$_2$)
sample listed in Table \ref{tab:sample-dimensions}. The large and
small graphite samples were used to measure the background from
neutron scattering on carbon in the CD$_2$ and CH$_2$ samples,
respectively. In addition to the samples listed in Table
\ref{tab:sample-dimensions}, empty target holders were used to measure
backgrounds from air scattering.

\begin{table}[h]
  \begin{ruledtabular}
    \caption{\label{tab:sample-dimensions}Properties of the scattering
      samples used.}
\centering
\begin{tabular}{lrrr}
  Sample & Mass (g) & Diameter (mm) & Height (mm)\\
  \hline
  CD$_{2}$         &  25.172  &  28.3  & 36.4 \\
  Large graphite  &  42.055  &  28.6  & 38.0 \\
  CH$_{2}$         &   3.389  &  14.2  & 22.7 \\
  Small graphite  &   2.924  &  9.4   & 23.6 \\
\end{tabular}
\end{ruledtabular}
\end{table}

The energies of the detected neutrons were determined from TOF
measurements. The incident neutron beam was pulsed at a repetition
rate of 2.5 MHz, and the width of each neutron bunch incident on the
scatterer was about 2 ns FWHM. The arrival of the deuteron beam pulse
on the neutron production gas cell was sensed with a capacitive beam
pickoff unit. A delayed signal derived from the beam pickoff unit was
used as the time reference for measuring the TOF of each detected
neutron. Pulse-shape discrimination techniques were used to reduce
backgrounds from gamma rays. A detector pulse-height threshold of
238.5 keVee ($\frac{1}{2} \times^{137}$Cs Compton edge) was applied,
where ``keVee'' denotes ``keV electron equivalent''.

\begin{figure}[htb]
\centering
\includegraphics[width=.9\linewidth]{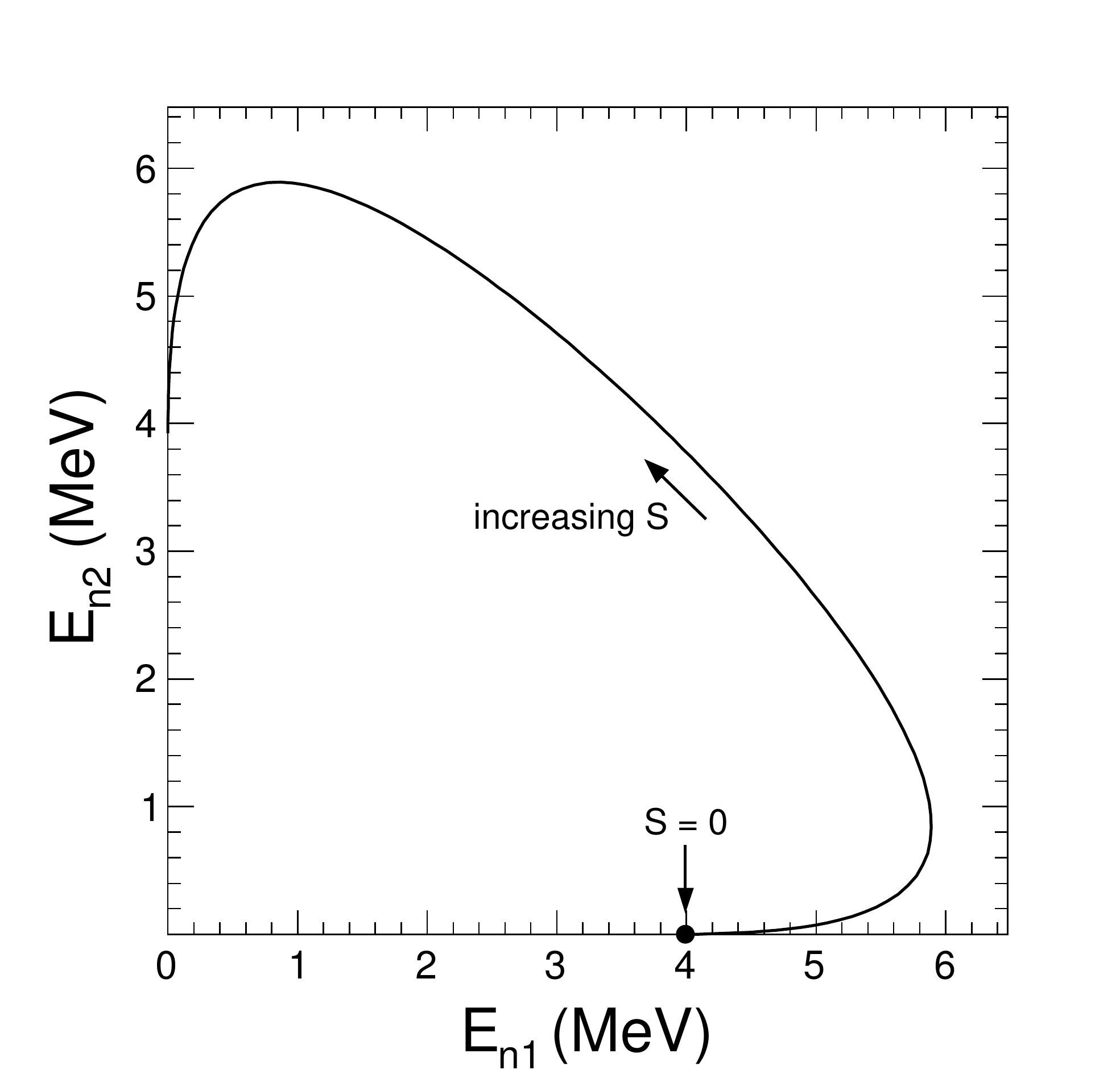}
\caption{\label{fig:locus} Plot of the kinematic locus of allowed
  neutron energies in \textit{nd} breakup for the central geometry of
  our experiment (see Fig. \ref{fig:ntof-setup}). The variable
  \textit{S} measures the arc length along the locus in the
  counterclockwise direction starting from the point where $E_{n2}$ is
  zero.}
\end{figure}

For the neutron elastic scattering measurements, a TOF histogram was
accumulated for the neutrons that were independently detected in each
of the two shielded detectors. Events from the \textit{nd} breakup
reaction were identified by the coincidence detection of neutrons in
the two shielded detectors. The \textit{nd} breakup events were
accumulated in a two-dimensional histogram of the TOF of the neutrons
detected in the left detector (D$_1$) versus the TOF of those detected
in the right detector (D$_2$). The events corresponding to the
\textit{nd} breakup reaction lie along a contour defined by the
reaction kinematics, i.e., the kinematic locus of the reaction or the
\textit{S} curve, as shown in Fig. \ref{fig:locus}. Arc length along
the kinematic locus is denoted by the variable \textit{S} and measured
in the counterclockwise direction starting at the point where the
energy of the second neutron is a minimum \cite{Glo96}.

A 100 ns wide time window was used to form the coincidences between
the signals from the two neutron detectors. The accidental coincidence
background was measured by forming coincidences between detector
signals caused by neutrons in two consecutive beam pulses. This was
achieved by delaying the signal from one detector by 400 ns, which is
one beam pulse period. With this technique, we detected and
accumulated TOF spectra for two categories of events: (1) an admixture
of \textit{nd}-breakup events and accidental-coincidence events, and
(2) purely accidental-coincidence events.

\section{Data Analysis}
\label{sec:analysis}

The differential cross section along the \textit{S} curve for the
\textit{nd} breakup reaction averaged over the kinematic acceptance of
our experiment setup was determined from the measured \textit{nn}
coincidence yields by Eq. \ref{eq:bu-cross-section}:
\begin{equation}
\label{eq:bu-cross-section}
\frac{d\sigma(\theta_1,\theta_2,\Delta\phi)}{ d\Omega_{1} d\Omega_{2} dS} =
\frac{Y_{nn}}{ \epsilon_{1}  \, \epsilon_{2} \,
  \alpha_{0}  \, \alpha_{1} \, \alpha_{2} \,
  N_{n} \, \rho_{D} \,
  d\Omega_{1}  \, d\Omega_{2} \, dS }  .
\end{equation}
The parameters in Eq. \ref{eq:bu-cross-section} are: the net
number of detected \textit{nn} coincidence events ($Y_{nn}$); the
efficiencies of the neutron detectors ($\epsilon_1$, $\epsilon_2$);
the transmission of the incident neutrons to the center of the CD$_2$
sample ($\alpha_0$); the transmission of the emitted neutrons through
the CD$_2$ sample and air to the face of each neutron detector
($\alpha_1$, $\alpha_2$); the number of neutrons incident on the
CD$_2$ sample ($N_n$); the nuclear areal density of the deuterium
sample ($\rho_{D}$ in nuclei/cm$^2$); the solid angles of the neutron
detectors ($d\Omega_1$, $d\Omega_2$); and the bin width along the
\textit{S} curve ($dS$). The scattering angles $\theta_1$ and $\theta_2$
are defined by the line that connects the center of the CD$_2$
scatterer to the center of each neutron detector, D$_1$ and D$_2$,
respectively, shown in Fig. \ref{fig:ntof-setup}. The azimuthal
opening angle $\Delta\phi$ is defined by the planes containing the
centers of D$_1$ and D$_2$ and the incident neutron beam
axis. Detector solid angles were calculated from the detector
radii and distances from the sample to the detectors, assuming a point
geometry. The Monte-Carlo simulation confirmed this assumption is
accurate to within 0.2\%.

\subsection{Determination of Breakup Yields}

\begin{figure}[htb]
\centering
\includegraphics[width=.9\linewidth]{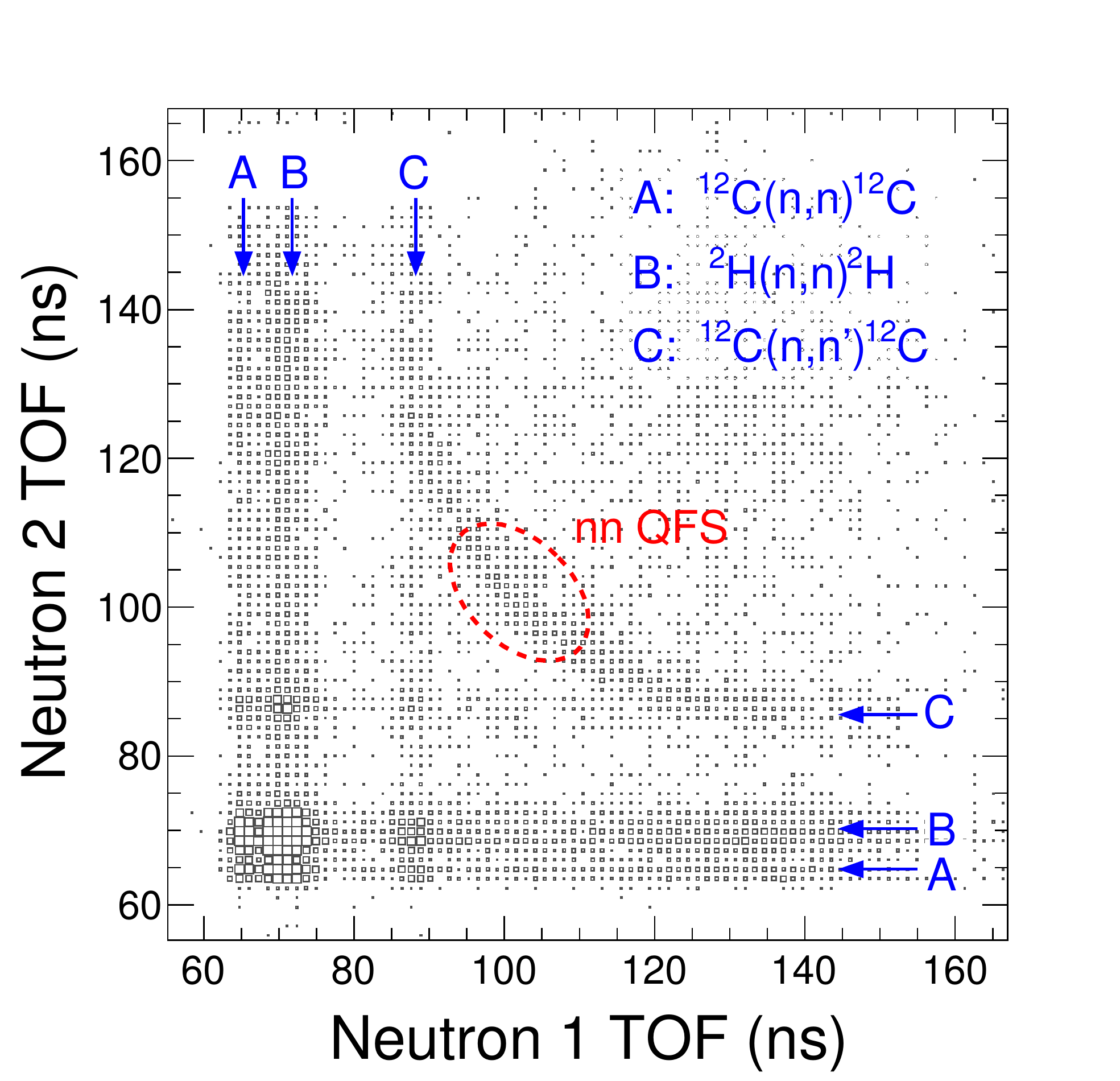}
\caption{\label{fig:ndbu-2d-coinc-spectrum}Raw two-dimensional neutron
  TOF coincidence spectrum accumulated with the setup shown in
  Fig. \ref{fig:ntof-setup} and described in Sec. \ref{sec:setup}. The
  vertical scale (i.e., the z-axis) is from a minimum of 1 count to a
  maximum of 50 counts per bin. The kinematic locus is clearly visible
  with the \textit{nn} QFS region circled by the red dashed curve. The
  main backgrounds from accidental coincidences are labeled by the
  blue arrows. This histogram was accumulated in 178 hours of data
  collection.}
\end{figure}

A raw two-dimensional coincidence neutron TOF spectrum is shown in
Fig. \ref{fig:ndbu-2d-coinc-spectrum}. The kinematic locus is clearly
visible, and the \textit{nn} QFS region at the center of the locus
(enclosed by the red dashed ellipse) is well separated from
backgrounds. Accidental coincidences due to elastic scattering from
deuterium and carbon and inelastic scattering from the first excited
state in carbon form bands parallel to the TOF axes; these are
identified by the labels A, B and C in
Fig. \ref{fig:ndbu-2d-coinc-spectrum}. The accidental coincidences
above and to the right of the kinematic locus are due to coincidences
between neutrons from \textit{nd} breakup events in which only one
neutron is detected and the elastic scattering of the continuum of
neutrons produced via deuteron breakup reactions in the neutron
production cell.

\begin{figure}[!hbt]
\centering
\includegraphics[width=.9\linewidth]{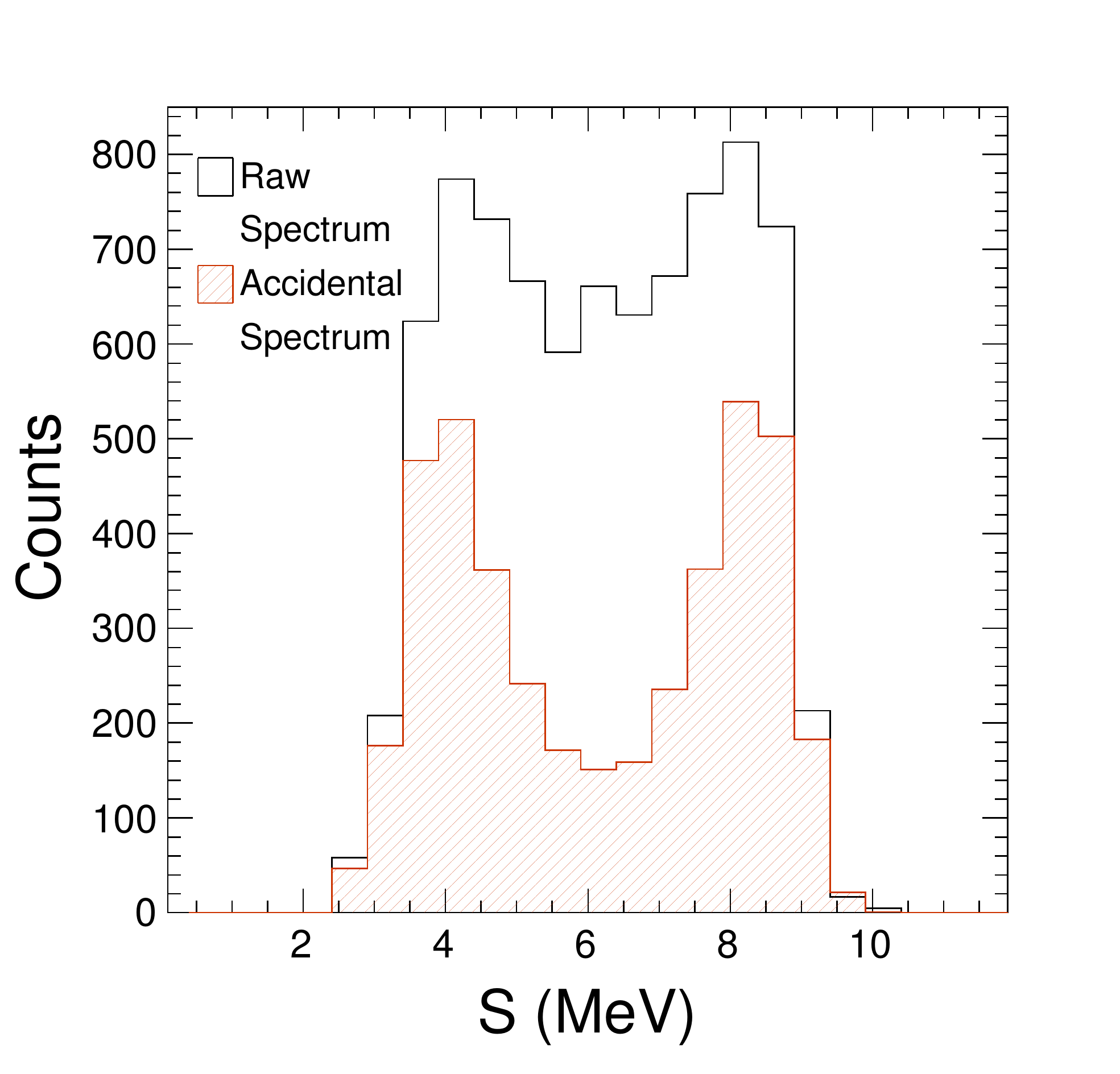}
\caption{\label{fig:countsAlongS}Raw and accidental neutron
  coincidence counts projected onto the \textit{S} curve. This
  histogram was accumulated in 577 hours of data collection. }
\end{figure}

Events in a band around the ideal point-geometry kinematic locus
(\textit{S} curve) defined by the central scattering angles of the
experiment $\theta_1, \theta_2,$ and $\Delta \phi$ were projected into
bins along the locus. The width of the band was determined by the
energy spread and angular acceptance of the experiment. Events were
projected using the method of Finckh \textit{et al.} \cite{Fin87}. The
\textit{S} curve was discretized in steps of 50 keV and each event was
projected to the closest point on the locus. Every event can be
represented by a point ($k^{exp}_{n1}, k^{exp}_{n2}$) in the
$k_{n1}-k_{n2}$ momentum plane, where $k_n$ is the momentum of a
neutron in the laboratory frame. Also, any point along the \textit{S}
curve can be represented in momentum space as ($k^{ideal}_{n1},
k^{ideal}_{n2}$). For each event, the squared distance in momentum
space between the event and every point on the \textit{S} curve
was calculated:
\begin{equation}
  \label{eq:event-projection}
  d^2 = \left( k_{n1}^{ideal} - k_{n1}^{exp} \right)^2 +
    \left( k_{n2}^{ideal} - k_{n2}^{exp} \right)^2 .
\end{equation}
For each event, the bin on the \textit{S} curve corresponding to the
minimum value of $d^2$ was incremented by one count. After
projecting onto the \textit{S} curve the yields were rebinned in
0.5-MeV-wide bins. The accidental coincidence data were analyzed in
the same way as the data that included the true detector coincidences
due to \textit{nd} breakup. The net \textit{nd} breakup yields were
computed bin-by-bin along the \textit{S} curve by subtracting the
accidental coincidence counts from the raw spectrum in each bin, see
Fig. \ref{fig:countsAlongS}. The cross section was determined from the
net coincidence yields in each bin along the \textit{S} curve.

\subsection{Detector Efficiency Measurements}
\label{sec:efficiency}

\begin{figure}[!hbt]
\centering
\includegraphics[width=.9\linewidth]{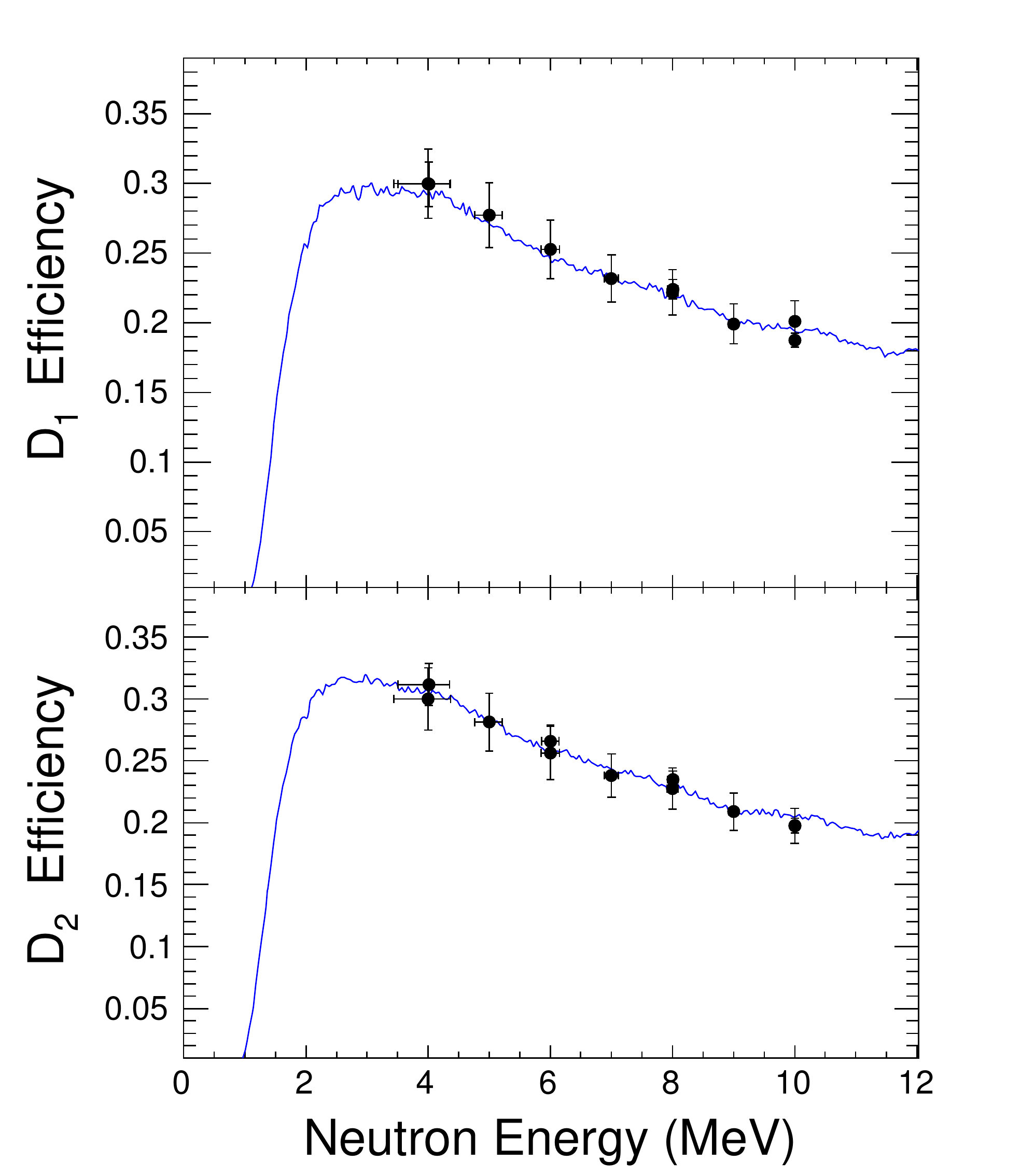}
\caption{\label{fig:detector-efficiencies} \textbf{Top}: Plot of
  efficiency for $D_1$. \textbf{Bottom}: Plot of efficiency for
  $D_2$. All efficiencies shown are for a pulse-height threshold of
  238.5 keVee ($\frac{1}{2} \times^{137}$Cs Compton edge). Measured
  efficiencies are indicated by the points. The vertical error bars
  include statistical and systematic uncertainties. The horizontal
  error bars show the calculated full energy spread due to deuteron
  energy loss in the gas cell used to produce the neutrons. Simulated
  detector efficiencies are shown by the curves. The simulation
  results for $D_1$ and $D_{2}$ have been scaled by 1.009 and 1.005 to
  fit the data. }
\end{figure}

Detector efficiencies were determined in a separate experiment by
measuring the neutron yield from the ${^2}\text{H(}d,n
\text{)}{^3}\text{He}$ reaction at zero degrees
\cite{Dro15}. Measurements were taken for neutron energies between 4
and 10 MeV in 1 MeV steps. The detector efficiency curves were
simulated using the code \textsc{neff7} \cite{Die82} between neutron
energies of 0 and 20 MeV in 50 keV steps. The results of the
\textsc{neff7} simulation were scaled to fit the measured
efficiencies, as shown in Fig. \ref{fig:detector-efficiencies}. The
simulated efficiencies agreed well with the data; the efficiency
curves for $D_1$ and $D_2$ were scaled up by 0.9\% and 0.5\%,
respectively, to fit the measured efficiencies. The scaled efficiency
curves were used in the Monte-Carlo simulation (see Sec. \ref{sec:monte-carlo}). 

\begin{figure}[!hbt]
  \centering
  \includegraphics[width=.9\linewidth]{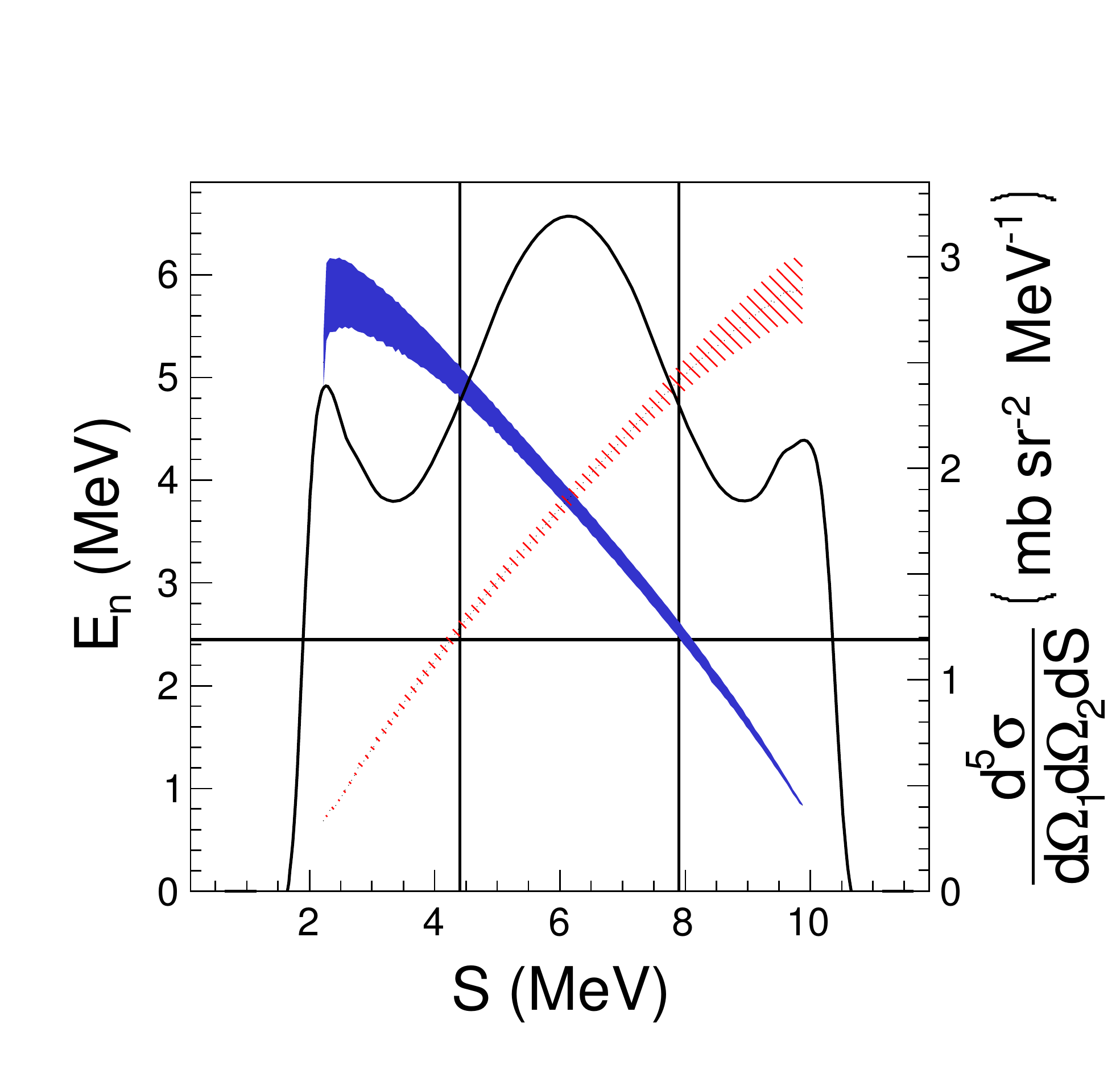}
  \caption{\label{fig:energy-along-S} Plot of the simulated neutron
    energies as a function of \textit{S} for \textit{nn} QFS in
    \textit{nd} breakup at 10.0 MeV. The detector setup is shown in
    Fig. \ref{fig:ntof-setup}. The energy of the neutrons is given on
    the left vertical axis. The bands show the energy spread (one
    standard deviation) about the average neutron energy in each
    bin. The solid blue band is the energy of the first neutron and
    the hashed red band is the energy of the second neutron. Also, the
    finite-geometry averaged cross section as a function of \textit{S}
    is shown by the solid black curve. The cross-section values are
    given on the vertical scale on the right side of the plot. The
    horizontal line shows the energy threshold of 2.45 MeV and the
    vertical lines show the region of the data passing this threshold
    cut.}
\end{figure}

At each end of the \textit{S} curve in the \textit{nn} QFS
configuration, one of the breakup neutrons has a very low energy. The
simulated energy of each neutron as a function of \textit{S} is
plotted in Fig. \ref{fig:energy-along-S} for our experiment setup. The
bands represent the energy spread of neutrons projected into each bin
along the \textit{S} curve (one standard deviation). As shown in
Fig. \ref{fig:detector-efficiencies}, the efficiency curves of the
neutron detectors rise sharply from the threshold energy of about 1
MeV up to about 2.3 MeV where the slope of the efficiency curve starts
to flatten as a function of neutron energy. Because the uncertainty in
the detector efficiency is greater than $\pm50\%$ near the threshold
energy, events that have a neutron with an energy of less than 2.45
MeV were rejected. The energy threshold cut is indicated by the
horizontal line in Fig. \ref{fig:energy-along-S}. This cut selects the
\textit{S}-curve region from 4.4 to 7.9 MeV for reporting
cross-section data, as indicated by the vertical lines in
Fig. \ref{fig:energy-along-S}.

\subsection{Monte-Carlo Simulation}
\label{sec:monte-carlo}

A Monte-Carlo (MC) simulation of the experiment was developed for four
purposes: (1) to allow for direct comparisons between the experiment
and theory by averaging the theoretical point-geometry cross sections
over the finite geometry and energy resolution of the experiment; (2)
to determine the average neutron transmission factors and detector
efficiencies used to convert the measured coincidence yields into a
cross section (see Eq. \ref{eq:bu-cross-section}); (3) to determine
quantitatively the effects of multiple scattering of neutrons in the
target; and (4) to quantify sources of background relevant to
extracting the \textit{nd} elastic yields.

\begin{figure}[htb]
\centering
\includegraphics[width=.9\linewidth]{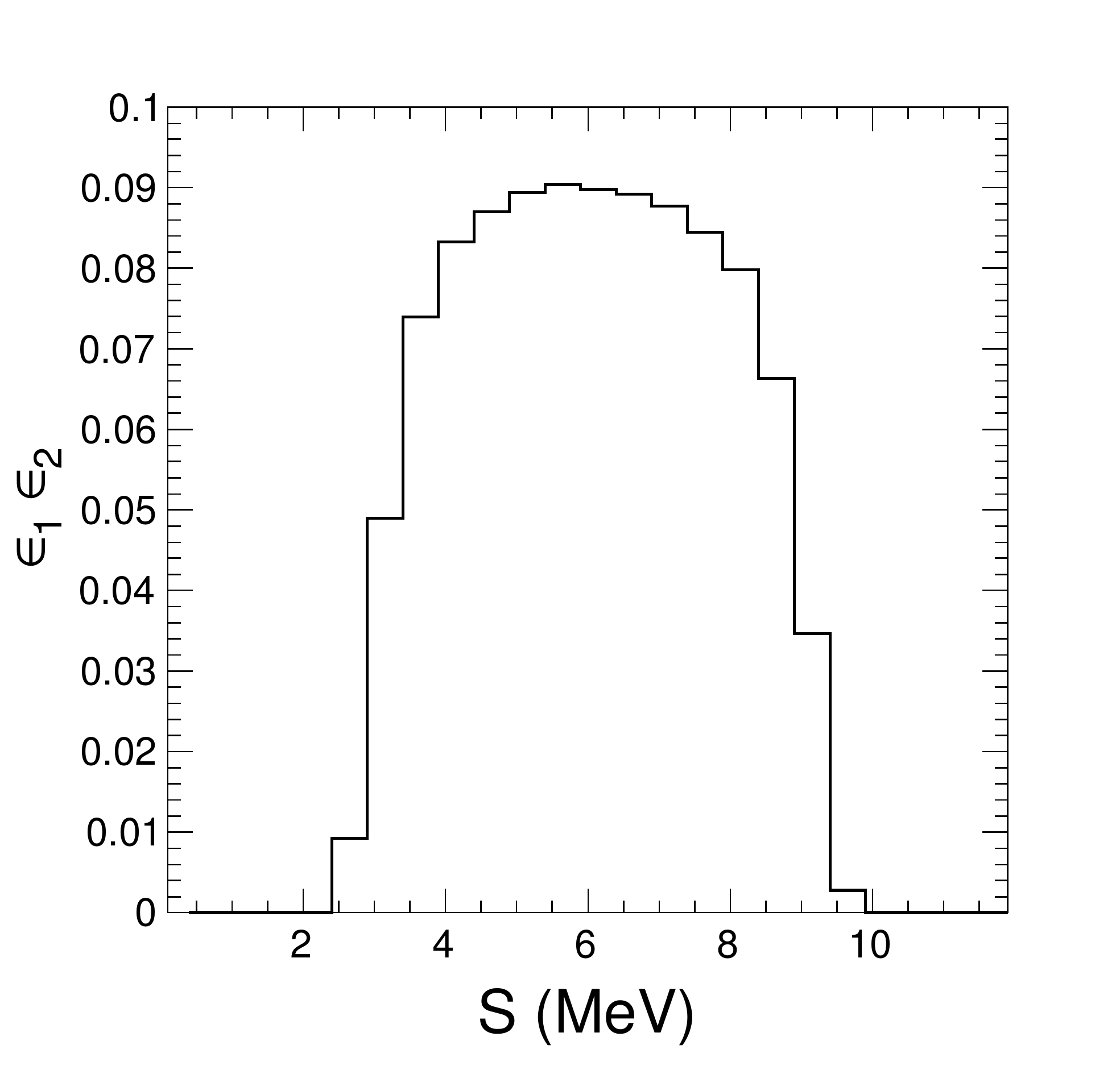}
\caption{\label{fig:avg-simulated-efficiency} Plot of the product of
  detector efficiencies $\epsilon_1 \epsilon_2$ as a function of
  \textit{S} averaged over the experiment geometry and energy
  spread using the MC simulation described in the text.}
\end{figure}

\begin{figure}[htb]
\centering
\includegraphics[width=.9\linewidth]{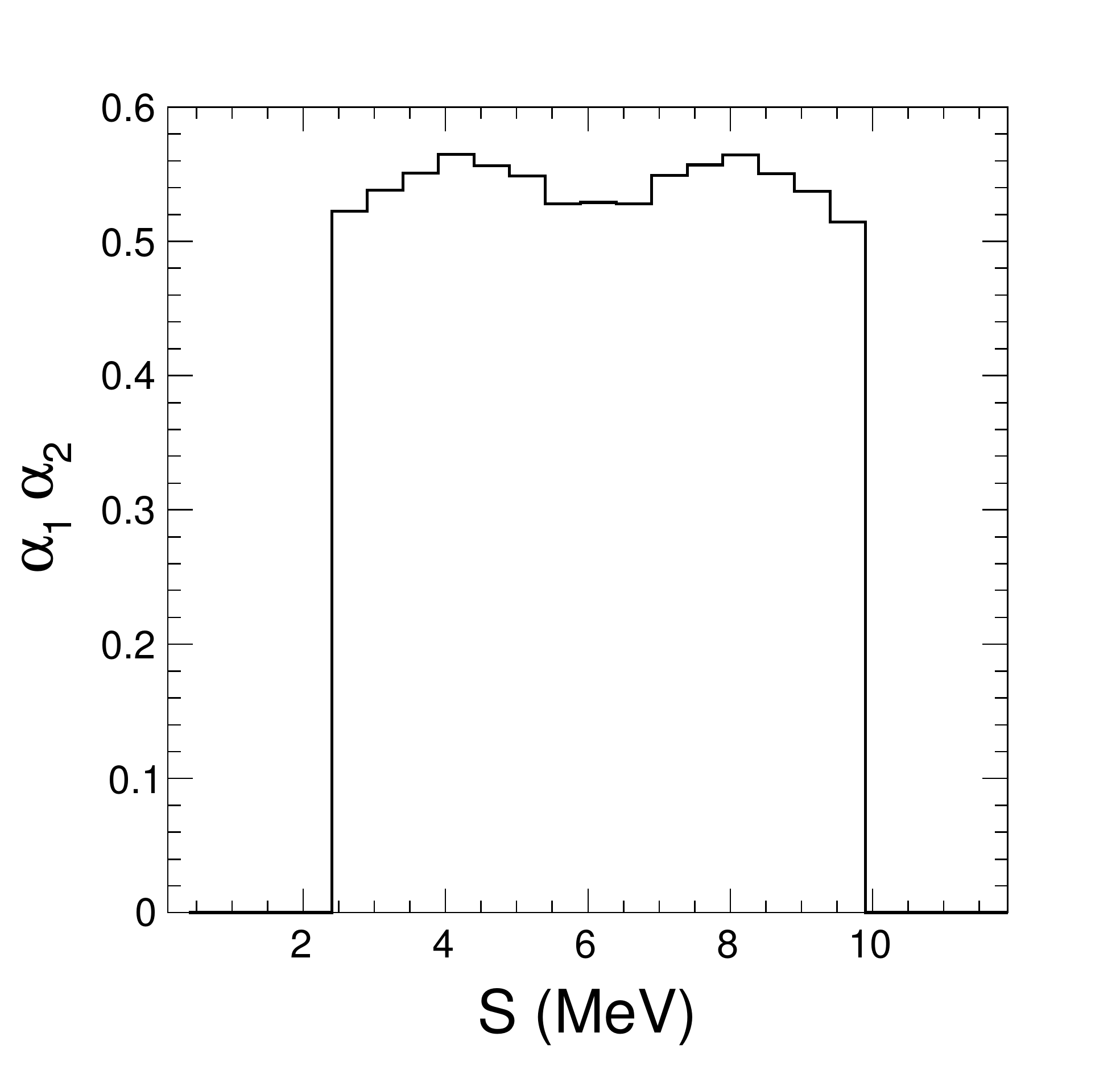}
\caption{\label{fig:avg-simulated-transmission} Plot of the product of
  neutron transmission factors $\alpha_1 \alpha_2$ as a function of
  \textit{S} averaged over the experiment geometry and energy spread
  using the MC simulation described in the text.}
\end{figure}

The MC simulation was used to average the breakup cross section over
the finite geometry of the experiment and to determine the average
detector efficiencies and transmission factors in
Eq. \ref{eq:bu-cross-section}. Scattering events were simulated by
tracing individual neutrons from their origin in the gas cell to the
detection of one or two neutrons in the liquid scintillators. A forced
scattering routine was used for computational efficiency. Details of
the MC simulation are described in the Appendix. Theoretical
point-geometry \textit{nd} breakup cross sections used in the
simulation were calculated by solving the three-body Faddeev equations
\cite{Fad61} with the CD-Bonn \textit{NN} potential \cite{Mac96} using
the technique described by Gl{\"o}ckle \textit{et al.}
\cite{Glo96}. Neutron detector efficiencies were determined using the
efficiency curves calculated with the code \textsc{neff7} as discussed
in Sec. \ref{sec:efficiency}. Finite-geometry averaged values for the
product of detector efficiencies $\epsilon_1 \epsilon_2$ as a function
of \textit{S} are shown in
Fig. \ref{fig:avg-simulated-efficiency}. Neutron transmission factors
were calculated using total neutron scattering cross sections from the
ENDF/B-VII.1 database \cite{Cha11}. Finite-geometry averaged values
for the product of neutron transmission factors $\alpha_1 \alpha_2$ as
a function of \textit{S} are shown in
Fig. \ref{fig:avg-simulated-transmission}.

Elastic scattering processes were also simulated for all four
scattering samples (see Table \ref{tab:sample-dimensions}). The
elastic scattering simulation used the same input data as the
\textit{nd} breakup simulation for detector efficiencies and neutron
transmission calculations. Cross sections for \textit{nd} elastic
scattering were calculated using the CD-Bonn \textit{NN}
potential. Cross sections for \textit{np} scattering were obtained
from the program \textsc{said} using the Bonn potential
\cite{SAID}. Cross sections for elastic and inelastic neutron
scattering from carbon were taken from Refs. \cite{Gla76,Cha11}.

\begin{figure}[htb]
\centering
\includegraphics[width=.9\linewidth]{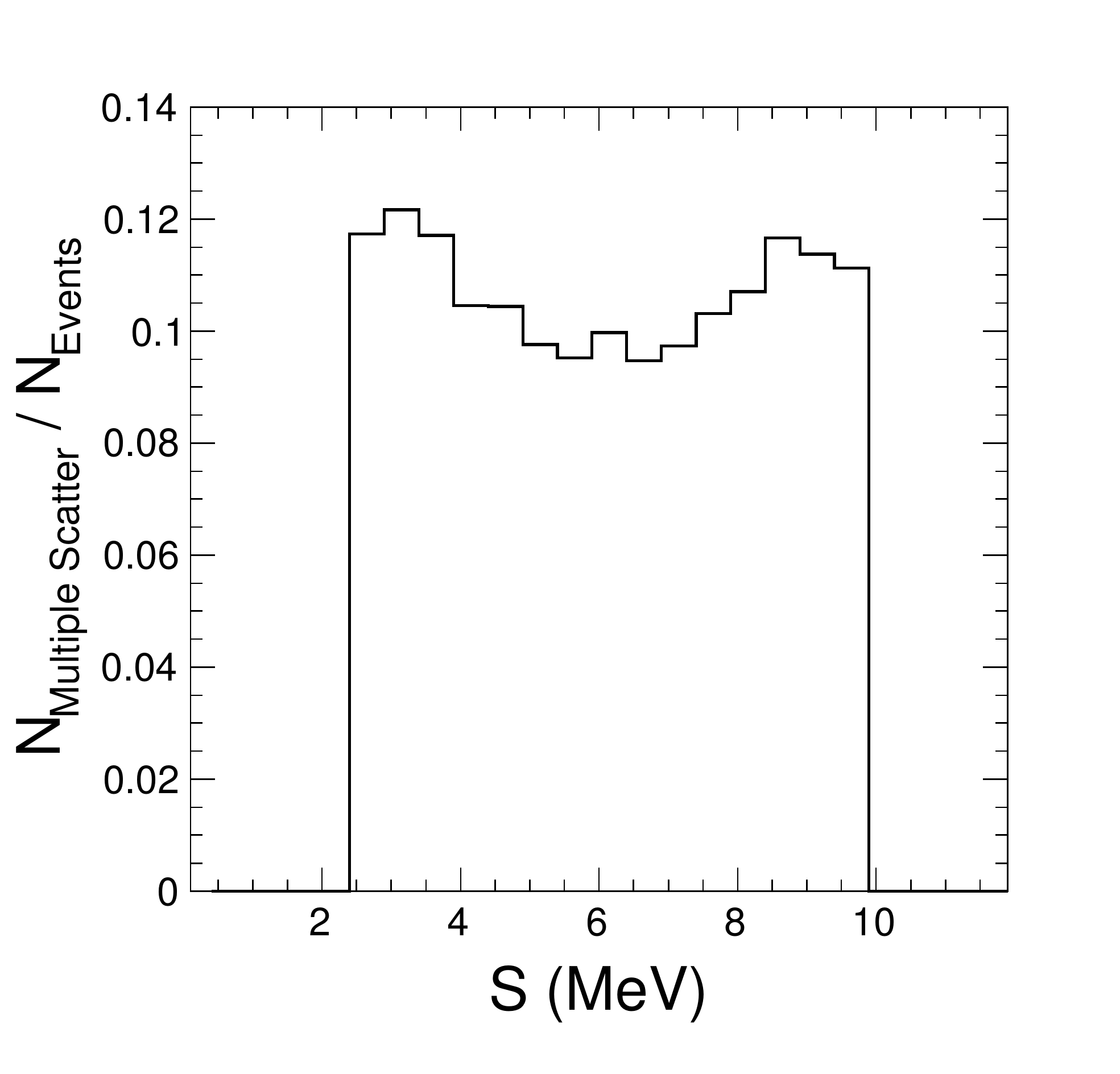}
\caption{\label{fig:bu-ms-frac} Plot of the fraction of simulated
  breakup events in which a neutron scattered twice as a function of
  \textit{S}.}
\end{figure}

The simulation was also used to study the effect of multiple
scattering of neutrons in the target on the extraction of \textit{nd}
breakup and elastic scattering yields from the measured neutron TOF
spectra. It was found that multiple scattering accounts for
about 9.9\% of the breakup yields near the QFS peak (see
Fig. \ref{fig:bu-ms-frac}) and only 5.0\% of the total yields in the
\textit{nd} elastic scattering peak (see
Fig. \ref{fig:nd-elastic-backgrounds}). In both cases, the measured
yields were corrected to account for multiple scattering.

Significant background was due to reactions induced with neutrons
produced by the ${^2}\text{H(}d,n \text{)}{^3}\text{He}$ reaction on
deuterons implanted in the tantalum beam stop at the end of the
neutron production gas cell. Simulations revealed neutrons produced in
the beam stop make up less than 0.1\% of the \textit{nn} coincidence
yields and about 2.7\% of the counts in the \textit{nd} elastic
scattering peak, as shown in Fig. \ref{fig:nd-elastic-backgrounds}.

As shown in Fig. \ref{fig:nd-elastic-backgrounds}, there is a small
background in the region of the \textit{nd} elastic TOF peak due to
neutron scattering from protons in the approximately 1.6\% CH$_2$
contaminant in the CD$_2$ sample. Because of the mass difference in
hydrogen and deuterium, less than half of these events fall within the
window of the \textit{nd} elastic TOF peak. Overall, the simulations
indicate that the \textit{np} scattering events contribue 0.8\% of the
total yields in the \textit{nd} elastic scattering window.

Another background quantified by the MC simulation was \textit{nd}
breakup events for which only one neutron was detected. As shown in
Fig. \ref{fig:nd-elastic-backgrounds}, the energy reach of neutrons
from the non-coincident breakup events is insufficient to contribute
to the yields in the window for the elastic TOF peak. These events do
contribute significantly to the background at long times in the TOF
spectra measured with the CD$_2$ sample. However, no such events are
present in TOF spectra measured with the graphite sample. This must be
carefully understood to ensure proper normalization of TOF spectra for
the graphite sample (see Sec. \ref{sec:luminosity}).

\subsection{Luminosity Determination}
\label{sec:luminosity}
The product of $N_n$ and $\rho_D$ in
Eq. \ref{eq:bu-cross-section} was determined from the yields
for \textit{nd} elastic scattering, which were measured concurrently
with the \textit{nd} breakup \textit{nn} coincidence yields. The
integrated beam-target luminosity is given by:
\begin{equation}
  \label{eq:luminosity}
  N_{n} \, \rho_{D} =
  \frac{Y_{nd}} { \epsilon_{nd} \, \alpha_{0} \,
    \alpha_{nd} \frac{d\sigma}{d\Omega} \, d\Omega } .
\end{equation}
The parameters in Eq. \ref{eq:luminosity} are: the net yields for
\textit{nd} elastic scattering ($Y_{nd}$); the efficiency of the
neutron detector at the energy of neutrons from \textit{nd} elastic
scattering ($\epsilon_{nd}$); the transmission of the incident
neutrons to the center of the sample ($\alpha_{0}$); the transmission
of the scattered neutrons through the sample and air to the
face of the neutron detector ($\alpha_{nd}$); the differential
scattering cross section for \textit{nd} elastic scattering
($\frac{d\sigma}{d\Omega}$); and the solid angle of the neutron
detector ($d\Omega$).

\begin{figure}[htb]
\centering
\includegraphics[width=.9\linewidth]{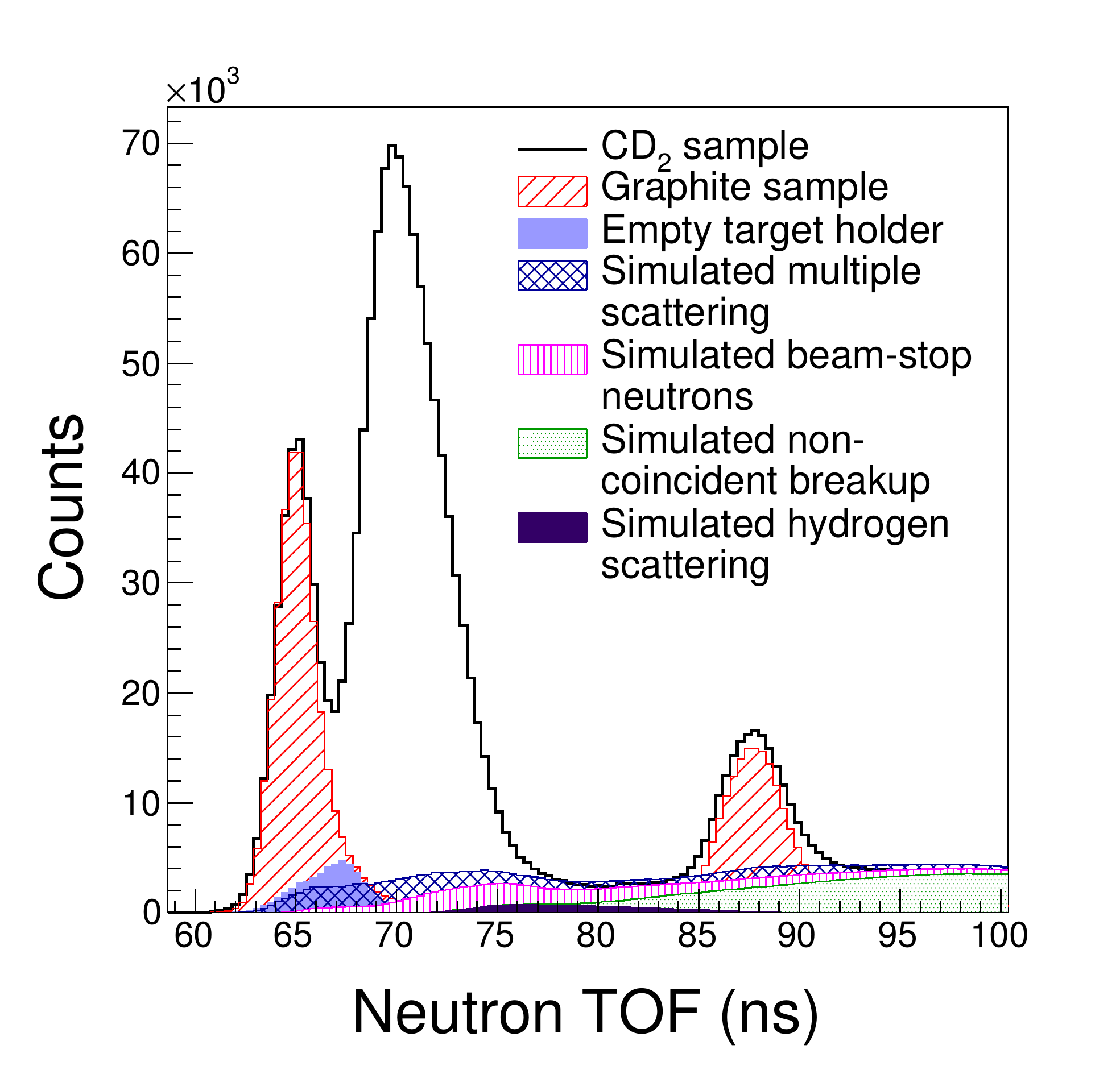}
\caption{\label{fig:nd-elastic-backgrounds} Plots of measured TOF
  spectra for scattering of 10.0 MeV neutrons from the CD$_2$ sample,
  graphite sample, and empty target holder at $\theta =
  36.7^{\circ}$. From left to right, the peaks in the spectrum are
  from elastic scattering on carbon, deuterium, and inelastic
  scattering from carbon. The plots include an overlay of the sum of
  simulated TOF spectra for multiple scattering of neutrons in the
  target, scattering of neutrons produced in the beam stop, neutrons
  scattering from hydrogen in the sample, and neutrons from
  non-coincident \textit{nd} breakup events. }
\end{figure}

An accurate extraction of the \textit{nd} elastic scattering yields
requires a detailed understanding of the backgrounds in the region of
the \textit{nd} elastic scattering peak in the neutron TOF spectrum as
shown in Fig. \ref{fig:nd-elastic-backgrounds}. Two major sources of
background were neutrons scattering from air and neutrons scattering
elastically from carbon. Scattering from air was measured using an
empty target holder and the background due to carbon was measured
using a graphite sample. The TOF spectra measured with the various
samples were normalized to each other using the integrated beam
current, the data acquisition system live time, and the gas pressure
in the neutron production cell. The empty sample TOF spectrum was
subtracted from the spectra measured with the CD$_2$ and carbon
samples.

The yields in the inelastic carbon scattering peak were used to finely
adjust the normalization factor of the spectrum obtained with the
graphite sample to the spectrum measured with the CD$_2$ sample. The
backgrounds due to neutron multiple scattering in the CD$_2$ sample,
non-coincident \textit{nd} breakup events, neutrons scattering from
hydrogen in the CD$_2$ sample, and neutrons produced via the
${^2}\text{H(}d,n\text{)}{^3}\text{He}$ reaction on deuterons
implanted in the beam stop were calculated using the MC simulation and
subtracted from the measured spectra.

We measured an integrated beam-target luminosity of $[4.41 \pm 0.0004
  \, \text{(stat)} \pm 0.2 \, \text{(sys)} ] \times 10^{36}$ cm$^{-2}$
in the left detector and $[ 4.40 \pm 0.0006 \, \text{(stat)} \pm 0.2
  \, \text{(sys)} ] \times 10^{36}$ cm$^{-2}$ in the right
detector. The average of these values was used in
Eq. \ref{eq:bu-cross-section} to calculate the breakup cross section:
\begin{equation}
  \label{eq:luminosity-geometric-mean}
  \langle N_n \rho_D \rangle = \sqrt{ (N_n \rho_D)_1 (N_n \rho_D)_2 }.
\end{equation}
A geometric mean was chosen to better cancel systematic uncertainties
in the final result. The value of $\langle N_n \rho_D \rangle$ used to
calculate the breakup cross section was $ [ 4.41 \pm 0.0004 \,
  \text{(stat)} \pm 0.2 \, \text{(sys)} ] \times 10^{36}$ cm$^{-2}$.

Sources of systematic uncertainty in the luminosity determination are
listed in Table \ref{tab:luminosity-systematics}. Uncertainties in the
yields for \textit{nd} scattering are mainly due to background
subtraction errors. Uncertainty in the absolute detector efficiencies
is due primarily to the uncertainties in the number of deuterium
nuclei in the gas cell and the background subtraction in the
efficiency measurements, as well as the uncertainties in the evaluated
${^2}\text{H(}d,n\text{)}{^3}\text{He}$ reaction cross sections used
to calculate the efficiencies \cite{Dro15,Dro87}. The uncertainty in
the relative detector efficiency is based on the variance between the
simulated detector efficiency curves and measured efficiencies (see
Fig. \ref{fig:detector-efficiencies}). A significant contribution to
the uncertainty in the detector efficiency is due to drifts in the
detector threshold (or gain) over time. Uncertainties in neutron
transmission are due to uncertainties in the total cross section data
\cite{Cha11}. The uncertainty in the cross section for \textit{nd}
scattering comes from the differences in the values given by different
\textit{NN} potentials \cite{Ski18}. The uncertainty in the solid
angle is mainly due to measurement errors in the distances from
the sample to the detectors. The uncertainties for the neutron
transmission factors and the uncertainties for the absolute detector
efficiencies are correlated. They must be summed before adding in
quadrature with the other uncorrelated uncertainties. This is
accounted for in Table \ref{tab:luminosity-systematics}.

\begin{table}[h]
\begin{ruledtabular}
\caption{\label{tab:luminosity-systematics} Sources of systematic
  uncertainty in the measurement of the beam-target luminosity
  $N_{n} \rho_D$. See text for details.}
\centering
\begin{tabular}{lr}
  Source  & Magnitude (\%) \\
  \hline
  Yields in \textit{nd} elastic peak &  2.3 \\
  Absolute detector efficiency & 3.9 \\ 
  Relative detector efficiency  &  1.1 \\ 
  Detector gain drift  &   0.5 \\
  Neutron transmission &   1.1 \\
  Cross section for \textit{nd} elastic scattering & 1.5 \\
  Solid angle & 0.4 \\
  \hline
  Total &  5.1 \\
\end{tabular}
\end{ruledtabular}
\end{table}

As a benchmark on our method for determining the beam-target
luminosity, the \textit{nd} elastic scattering cross section was
determined relative to the \textit{np} scattering cross section at
32\textdegree{} in the lab. This angle was chosen to kinematically
separate neutrons scattering on hydrogen from neutrons scattering on
carbon. The \textit{np} scattering yields were extracted from
TOF spectra in the same way as the \textit{nd} elastic scattering
yields. The \textit{np} scattering cross sections used in this
work were obtained from Ref. \cite{SAID}.

We measured an \textit{nd} elastic scattering cross-section value of
$213.7 \pm 0.1 \, \text{(stat)} \pm 10.9 \, \text{(sys) mb/sr}$ in the
left detector and $216.0 \pm 0.1 \, \text{(stat)} \pm 11.0 \,
\text{(sys) mb/sr}$ in the right detector. This is in agreement with
the value of 205.0 mb/sr predicted by the finite-geometry theoretical
calculations using the CD-Bonn \textit{NN} potential.

\begin{table}[ht]
\begin{ruledtabular}
\caption{\label{tab:nd-elastic-systematics} Sources of systematic
  uncertainty in the measurement of the \textit{nd} elastic scattering
  cross section.}
\centering
\begin{tabular}{lr}
  Source  & Magnitude (\%) \\
  \hline
  Yields in \textit{nd} elastic peak & 3.2 \\ 
  Yields in \textit{np} peak & 1.4 \\
  Finite geometry correction & 2.2 \\
  Relative detector efficiency  &  2.1 \\ 
  Detector gain drift  &  1.4 \\ 
  Number of deuterium nuclei & 1.0 \\
  Number of hydrogen nuclei & 0.4 \\
  Neutron transmission & 0.8 \\
  Cross section for \textit{np} scattering & 0.4 \\
  Live time correction & 0.6 \\
  \hline
  Total &  5.1 \\
\end{tabular}
\end{ruledtabular}
\end{table}

The sources of systematic errors in our measurement of the \textit{nd}
elastic scattering cross section are listed in Table
\ref{tab:nd-elastic-systematics}. Uncertainties in the yields for
\textit{np} scattering are due to background subtraction errors. There
is substantial uncertainty in the finite-geometry correction used to
account for the difference in the average flux seen by the CD$_2$ and
CH$_2$ samples because of their sizes relative to their distance from
the neutron production gas cell. Drifts in the detector bias over time
change the detector efficiencies, resulting in a significant
uncertainty. The uncertainty in the number of nuclei in the CD$_2$
sample is due to the unknown chemical purity of the sample, which is
listed as $>98\%$. We have assumed the chemical purity is $100\%$ with
an error of $1\%$. There is little uncertainty in the isotopic
enrichment, measured by Cambridge Isotope Laboratories to be $98.4 \pm
0.1\%$. The uncertainty in the number of hydrogen nuclei in the CH$_2$
sample is mainly due to the uncertainty in the measured mass of that
sample. The uncertainty in the cross section for \textit{np}
scattering is the difference in the values given by different
\textit{NN} potentials and partial-wave analyses \cite{SAID}. The data
acquisition system live time was measured in two ways: (1) the number
of event triggers passing the data acquisition system (DAQ) veto were
compared to the total number of event triggers, and (2) the number of
pulses from a 60 Hz clock passing the DAQ veto were compared to the
total number of clock pulses. The DAQ veto was the logical \textsc{or}
of the analog-to-digital converter busy signal, time-to-digital converter busy signal, and DAQ computer readout signals. The live
time measured by the triggers was used to compute the \textit{nd}
elastic cross section, and the associated uncertainty is the
difference between the live times determined by the two methods. All
errors are uncorrelated and added in quadrature.

\section{Results}
\label{sec:results}

\begin{figure}[htb]
\centering
\includegraphics[width=.9\linewidth]{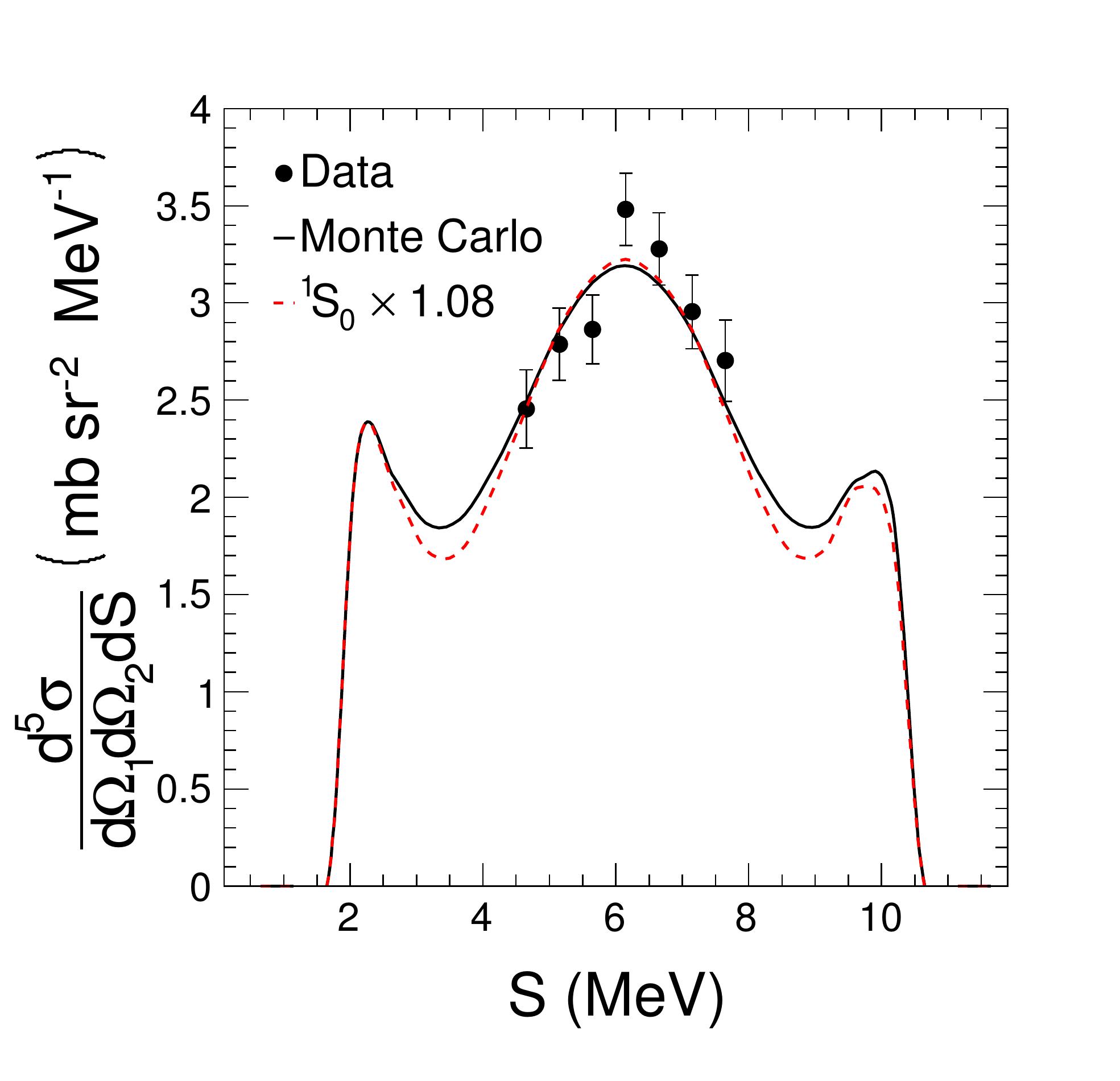}
\caption{\label{fig:nnqfs-xsec} A plot of the measured \textit{nn} QFS
  cross section (circles) and the result of the MC simulation (solid
  curve) for \textit{nd} breakup at 10.0 MeV. The experiment setup is
  shown in Fig. \ref{fig:ntof-setup}. The red dashed curve is the
  result of the MC simulation performed with the $^{1}S_{0}$
  \textit{nn} interaction matrix elements scaled by $\times
  1.08$. Error bars represent statistical uncertainties only; there is
  also a systematic uncertainty of $\pm5.6$\%.}
\end{figure}

Our cross-section data for \textit{nn} QFS in \textit{nd} breakup at
10.0 MeV are plotted as a function of \textit{S} in
Fig. \ref{fig:nnqfs-xsec}. The curves are predictions of rigorous $3N$
calculations based on the CD-Bonn potential that have been averaged
over the finite geometry and energy resolution of the experiment using
the MC simulation. The error bars on the data points represent only
statistical uncertainties; there is also a systematic uncertainty of
$\pm$5.6\%. The sources of systematic uncertainty are listed in Table
\ref{tab:nn-QFS-systematics} and summarized below.

The error in the coincidence yields is due to the uncertainty in the
correction for multiple scattering of neutrons in the sample. The
uncertainty due to detector gain drift is determined by the percent
change in the ratio of efficiencies at the energies for \textit{nd}
elastic scattering and \textit{nn} QFS breakup for small drifts in the
gain. All other sources of uncertainty are the same as those discussed
in Sec. \ref{sec:luminosity} for the determination of the integrated
beam-target luminosity. Combining
Eqs. \ref{eq:luminosity-geometric-mean}, \ref{eq:luminosity}, and
\ref{eq:bu-cross-section} leads to a reduction of several
uncertainties. Specifically, the neutron transmission factors
$\alpha_{0}$ for the incident neutron cancel, resulting in elimination
of that uncertainty. Uncertainties in the neutron transmission factors
and the absolute detector efficiencies for each detector are
correlated and must be added linearly with opposite signs for factors
in the numerator and denominator. The square root in
Eq. \ref{eq:luminosity-geometric-mean} results in a factor of $\approx
\frac{1}{\sqrt{2}}$ multiplying uncertainties associated with the
luminosity. Summing all uncertainties in quadrature with the
appropriate factors (accounted for in Table
\ref{tab:nn-QFS-systematics}) gives a total systematic uncertainty of
$\pm 5.6\%$.

\begin{table}[htb]
\begin{ruledtabular}
\caption{\label{tab:nn-QFS-systematics} Sources of systematic
  uncertainty in the measurement of the \textit{nn} QFS cross
  section. See text for details.}
\centering
\begin{tabular}{lr}
  Source  & Magnitude (\%) \\
  \hline
  Coincidence yields & 1.0 \\ 
  Absolute detector efficiency  &  3.9 \\ 
  Relative detector efficiency  &  2.4 \\ 
  Detector gain drift  &  1.1 \\ 
  Neutron transmission &  0.8 \\ 
  Yields in \textit{nd} elastic peak & 2.3 \\ 
  Cross section for \textit{nd} elastic scattering & 1.5 \\
  Solid angle & 0.4 \\
  \hline
  Total &  5.6 \\
\end{tabular}
\end{ruledtabular}
\end{table}

The data are in good agreement with the theoretical prediction
($\chi^2$ per datum = 0.97). Integrating the cross section from
\textit{S} = 4.4 MeV to 7.9 MeV gives an integrated measured cross
section of $ 20.5 \pm 0.5 \text{(stat)} \pm 1.1 \text{(sys)} $
mb/sr$^{2}$, which is consistent with the simulated value of 20.1
mb/sr$^{2}$. Scaling the \textit{nn} $^{1}S_{0}$ interaction by a
factor of $\times 1.08$ slightly increases the chi-squared value of the
comparison between data and theory ($\chi^2$ per datum = 0.98);
however, this change is not significant.

\section{Conclusions}
\label{sec:conclusion}

We have measured the \textit{nn} QFS cross section in \textit{nd}
breakup at an incident neutron beam energy of 10.0 MeV using standard
neutron TOF techniques. Our setup used a pulsed beam with an open
neutron source and heavily shielded neutron detectors. This was the
first measurement of this quantity using this detector and source
arrangement. The theoretical prediction agrees well with the data
within the uncertainty of the experiment. The good agreement between
our data and the $3N$ calculations indicates that the technique of
using \textit{nd} elastic scattering to determine the beam-target
luminosity works well for this type of measurement. With this method
we were able to determine the beam-target luminosity to an accuracy of
$\pm 5.1\%$. As expected, we are unable to rule out the validity of
the observed discrepancies between previously reported data
\cite{Sie02,Rua07,Lub92} and theory for the cross section for
\textit{nn} QFS in \textit{nd} breakup. The \textit{nn} QFS dilemma
remains unresolved, suggesting the possibility of significant
charge-symmetry breaking in the \textit{NN} system. New measurements
of \textit{nn} QFS in \textit{nd} breakup should be performed at
higher energies using a collimated neutron beam for maximum
sensitivity to the $^{1}S_{0}$ \textit{nn} interaction and using
\textit{nd} elastic scattering to determine the beam-target
luminosity, a technique validated in this work. The measurements
should be carried out with a systematic uncertainty less than
$\pm$4\%.

\begin{acknowledgments}
The authors thank the TUNL technical staff for their
contributions. The authors appreciate the use of the supercomputer
cluster of the JSC in J{\"u}lich, Germany, where part of the numerical
calculations were performed. This work is supported in part by the
U.S. Department of Energy under grant Nos. DE-FG02-97ER41033 and
DE-SC$0$005367 and by the Polish National Science Center under grant
No. DEC-2016/22/M/ST2/00173.
\end{acknowledgments}

\appendix*
\section{}
\label{sec:app}
Here we discuss the details of the \textit{nd} breakup MC simulation
and note that the elastic scattering simulation follows a similar
procedure. The steps for simulating a single \textit{nd} breakup
history are outlined below.

\begin{enumerate}
\item \label{step:select-points} A point was randomly selected in
  each: the neutron production cell, the scattering sample, and both
  detectors. These points fix the scattering angles $ \theta_1,
  \theta_2, $ and $\Delta \phi$ for the event. The incident neutron
  energy $E_0$ was calculated from the incident deuteron energy and
  the kinematics of the ${^2}\text{H(}d,n \text{)}{^3}\text{He}$
  reaction. The deuteron energy in the gas cell was approximated as a
  linear function of distance along the axis of the gas cell due to
  the energy loss in the gas by the deuterons.
\item An intensity factor for the incident neutrons, $I(E_0)$, was
  calculated from the ${^2}\text{H(}d,n \text{)}{^3}\text{He}$
  reaction cross section \cite{Dro15} and neutron transmission from
  the production point in the gas cell to the breakup point in the
  CD$_2$ sample.
\item The breakup cross section was determined in steps of 50 keV
  along the \textit{S} curve using a multiparameter interpolation over
  a library of theoretical point-geometry cross sections. The incident
  neutron energy $E_0$, the scattering angles $ \theta_1, \theta_2,
  \Delta \phi$, and the position along the \textit{S} curve were used
  as interpolation parameters.
\item The product of detector efficiencies $\epsilon_1 \epsilon_2$ and
  neutron transmission factors $\alpha_1 \alpha_2$ were calculated for
  points at 50 keV intervals along the \textit{S} curve of the
  simulated event.
\item \label{step:wgt} A weight factor $w(S)$ used to calculate
  the average breakup cross section was tabulated. The weight factor
  for each breakup event is:
  \begin{equation}
    \label{eq:mc-weight-factor}
    w_{i}(S) = I(E_{0}) \epsilon_{1} \epsilon_{2} \alpha_{1} \alpha_{2}.
  \end{equation}
\item The TOF of each neutron was computed at each point along the
  \textit{S} curve. The simulated TOF $t_{sim}$ was used with the
  center-to-center distance from the sample to detector $d'$ to
  calculate the energy $E'$ of each breakup neutron in the same way as
  the experiment:
  \begin{equation}
    \label{eq:measured-energy}
    E' = \frac{1}{2} m_{n} \left( \frac{d'}{t_{sim}} \right)^2 .
  \end{equation}
\item The energies $E'_{1}, E'_{2}$ of the two breakup neutrons were
  used to project each simulated event onto the point-geometry
  kinematic locus in the same way as the experimental data (see
  Eq. \ref{eq:event-projection}). For each simulated event, the weight
  factor, the values of the breakup cross section, the detector
  efficiencies, and the neutron transmission factors were stored in
  bins along the point-geometry \textit{S} curve.
\end{enumerate}

After simulating a sufficient number of histories (\mbox{$\sim
  10^6$}), the finite-geometry averaged values of the breakup cross
section, detector efficiencies, and neutron transmission factors as a
function of \textit{S}-curve length were calculated using the weight
factors from step \ref{step:wgt} above. The formula for calculating
the average breakup cross section is given by:
\begin{equation}
  \label{eq:mc-xsec}
  \left\langle \frac{d^5\sigma (S)}{ d\Omega_1 d\Omega_2 dS}
  \right\rangle_{MC} =
  \frac{  \sum_{i} w_{i}(S) \sigma_{i}(S)  }
       { \sum_{i} w_{i}(S) } ,
\end{equation}
where $\sigma_{i}(S)$ is the breakup cross section, $w_{i}(S)$ are given
by Eq. \ref{eq:mc-weight-factor}, and the index $i$ runs over
events. The average product of detector efficiencies and neutron
transmission factors are calculated similarly.

Theoretical point-geometry cross sections in the interpolation library
were calculated with the CD-Bonn \textit{NN} potential
\cite{Mac96}. The library was a five-dimensional array indexed by the
incoming neutron energy $E_{0}$, the scattering angles $\theta_1,
\theta_2$, and $\Delta \phi$, and the position along the \textit{S}
curve. The range of the library indices spanned all possible
scattering configurations for the geometry of our experiment and the
step size in each dimension was chosen to minimize the variance
between points on the grid while keeping the library to a reasonable
size ($\sim 10^6$ points).

Some modifications to the simulation procedure outlined above are
necessary to simulate various backgrounds. To simulate multiple
scattering, a second point was randomly chosen within the scattering
volume. Processes with more than two neutron scattering sites were not
considered. The simulation process was otherwise the same as described
above. In the case of elastic scattering, all permutations of
scattering from two nuclei in the sample were simulated.  For
\textit{nd} breakup, two cases were simulated: (1) elastic scattering
followed by \textit{nd} breakup, and (2) \textit{nd} breakup followed
by elastic scattering of one of the breakup neutrons. A second breakup
cross section library spanning all kinematically allowed breakup
configurations was generated for the simulation of multiple
scattering.

In the case of non-coincident \textit{nd} breakup events, a random
direction was chosen for one of the two breakup neutrons rather than a
point in a detector. Since the undetected neutron can be emitted in
any kinematically allowed direction, the same breakup
cross section library used for simulating multiple scattering was
used to simulate non-coincident breakup. Weight factors were
calculated using only the neutron transmission factor and detector
efficiency for the detected neutron.

To simulate neutrons produced via the
${^2}\text{H(}d,n\text{)}{^3}\text{He}$ on deuterons implanted in the
beam stop, the neutron production point was chosen inside the tantalum
beam stop. All scattering processes were then simulated in the normal
way. The distribution of deuterons implanted in the beam stop was
assumed to be uniform. Deuteron energy loss in the tantalum was
simulated using \textsc{srim} \cite{SRIM}. The ratio of deuterons in
the beam stop to deuterons in the gas was determined by comparing
simulated and measured TOF spectra for the monitor detector positioned
at 3\textdegree{} with no sample present. The number of deuterons in
the beam stop was determined to be 6.9\% of the number of deuterons in
the gas.


\begin{thebibliography}{22}%
\makeatletter
\providecommand \@ifxundefined [1]{%
 \@ifx{#1\undefined}
}%
\providecommand \@ifnum [1]{%
 \ifnum #1\expandafter \@firstoftwo
 \else \expandafter \@secondoftwo
 \fi
}%
\providecommand \@ifx [1]{%
 \ifx #1\expandafter \@firstoftwo
 \else \expandafter \@secondoftwo
 \fi
}%
\providecommand \natexlab [1]{#1}%
\providecommand \enquote  [1]{``#1''}%
\providecommand \bibnamefont  [1]{#1}%
\providecommand \bibfnamefont [1]{#1}%
\providecommand \citenamefont [1]{#1}%
\providecommand \href@noop [0]{\@secondoftwo}%
\providecommand \href [0]{\begingroup \@sanitize@url \@href}%
\providecommand \@href[1]{\@@startlink{#1}\@@href}%
\providecommand \@@href[1]{\endgroup#1\@@endlink}%
\providecommand \@sanitize@url [0]{\catcode `\\12\catcode `\$12\catcode
  `\&12\catcode `\#12\catcode `\^12\catcode `\_12\catcode `\%12\relax}%
\providecommand \@@startlink[1]{}%
\providecommand \@@endlink[0]{}%
\providecommand \url  [0]{\begingroup\@sanitize@url \@url }%
\providecommand \@url [1]{\endgroup\@href {#1}{\urlprefix }}%
\providecommand \urlprefix  [0]{URL }%
\providecommand \Eprint [0]{\href }%
\providecommand \doibase [0]{http://dx.doi.org/}%
\providecommand \selectlanguage [0]{\@gobble}%
\providecommand \bibinfo  [0]{\@secondoftwo}%
\providecommand \bibfield  [0]{\@secondoftwo}%
\providecommand \translation [1]{[#1]}%
\providecommand \BibitemOpen [0]{}%
\providecommand \bibitemStop [0]{}%
\providecommand \bibitemNoStop [0]{.\EOS\space}%
\providecommand \EOS [0]{\spacefactor3000\relax}%
\providecommand \BibitemShut  [1]{\csname bibitem#1\endcsname}%
\let\auto@bib@innerbib\@empty
\bibitem [{\citenamefont {Gl{\"o}ckle}\ \emph {et~al.}(1996)\citenamefont
  {Gl{\"o}ckle}, \citenamefont {Wita{\l}a}, \citenamefont {H{\"u}ber},
  \citenamefont {Kamada},\ and\ \citenamefont {Golak}}]{Glo96}%
  \BibitemOpen
  \bibfield  {author} {\bibinfo {author} {\bibfnamefont {W.}~\bibnamefont
  {Gl{\"o}ckle}}, \bibinfo {author} {\bibfnamefont {H.}~\bibnamefont
  {Wita{\l}a}}, \bibinfo {author} {\bibfnamefont {D.}~\bibnamefont
  {H{\"u}ber}}, \bibinfo {author} {\bibfnamefont {H.}~\bibnamefont {Kamada}}, \
  and\ \bibinfo {author} {\bibfnamefont {J.}~\bibnamefont {Golak}},\
  }\href@noop {} {\bibfield  {journal} {\bibinfo  {journal} {Phys. Rep.}\
  }\textbf {\bibinfo {volume} {274}},\ \bibinfo {pages} {107} (\bibinfo {year}
  {1996})}\BibitemShut {NoStop}%
\bibitem [{\citenamefont {Siepe}\ \emph {et~al.}(2002)\citenamefont {Siepe},
  \citenamefont {Deng}, \citenamefont {Huhn}, \citenamefont {W{\"a}tzold},
  \citenamefont {Weber}, \citenamefont {von Witsch}, \citenamefont
  {Wita{\l}a},\ and\ \citenamefont {Gl{\"o}ckle}}]{Sie02}%
  \BibitemOpen
  \bibfield  {author} {\bibinfo {author} {\bibfnamefont {A.}~\bibnamefont
  {Siepe}}, \bibinfo {author} {\bibfnamefont {J.}~\bibnamefont {Deng}},
  \bibinfo {author} {\bibfnamefont {V.}~\bibnamefont {Huhn}}, \bibinfo {author}
  {\bibfnamefont {L.}~\bibnamefont {W{\"a}tzold}}, \bibinfo {author}
  {\bibfnamefont {C.}~\bibnamefont {Weber}}, \bibinfo {author} {\bibfnamefont
  {W.}~\bibnamefont {von Witsch}}, \bibinfo {author} {\bibfnamefont
  {H.}~\bibnamefont {Wita{\l}a}}, \ and\ \bibinfo {author} {\bibfnamefont
  {W.}~\bibnamefont {Gl{\"o}ckle}},\ }\href@noop {} {\bibfield  {journal}
  {\bibinfo  {journal} {Phys. Rev. C}\ }\textbf {\bibinfo {volume} {65}},\
  \bibinfo {pages} {034010} (\bibinfo {year} {2002})}\BibitemShut {NoStop}%
\bibitem [{\citenamefont {Ruan}\ \emph {et~al.}(2007)\citenamefont {Ruan},
  \citenamefont {Zhou}, \citenamefont {Li}, \citenamefont {Jiang},
  \citenamefont {Huang}, \citenamefont {Zhong}, \citenamefont {Tang},
  \citenamefont {Qi}, \citenamefont {Bao}, \citenamefont {Xin}, \citenamefont
  {von Witsch},\ and\ \citenamefont {Wita{\l}a}}]{Rua07}%
  \BibitemOpen
  \bibfield  {author} {\bibinfo {author} {\bibfnamefont {X.~C.}\ \bibnamefont
  {Ruan}}, \bibinfo {author} {\bibfnamefont {Z.~Y.}\ \bibnamefont {Zhou}},
  \bibinfo {author} {\bibfnamefont {X.}~\bibnamefont {Li}}, \bibinfo {author}
  {\bibfnamefont {J.}~\bibnamefont {Jiang}}, \bibinfo {author} {\bibfnamefont
  {H.~X.}\ \bibnamefont {Huang}}, \bibinfo {author} {\bibfnamefont {Q.~P.}\
  \bibnamefont {Zhong}}, \bibinfo {author} {\bibfnamefont {H.~Q.}\ \bibnamefont
  {Tang}}, \bibinfo {author} {\bibfnamefont {B.~J.}\ \bibnamefont {Qi}},
  \bibinfo {author} {\bibfnamefont {J.}~\bibnamefont {Bao}}, \bibinfo {author}
  {\bibfnamefont {B.}~\bibnamefont {Xin}}, \bibinfo {author} {\bibfnamefont
  {W.}~\bibnamefont {von Witsch}}, \ and\ \bibinfo {author} {\bibfnamefont
  {H.}~\bibnamefont {Wita{\l}a}},\ }\href@noop {} {\bibfield  {journal}
  {\bibinfo  {journal} {Phys. Rev. C}\ }\textbf {\bibinfo {volume} {75}},\
  \bibinfo {pages} {057001} (\bibinfo {year} {2007})}\BibitemShut {NoStop}%
\bibitem [{\citenamefont {L{\"u}bcke}(1992)}]{Lub92}%
  \BibitemOpen
  \bibfield  {author} {\bibinfo {author} {\bibfnamefont {W.}~\bibnamefont
  {L{\"u}bcke}},\ }\href@noop {} {Ph.D. thesis},\ \bibinfo  {school}
  {University of Bochum} (\bibinfo {year} {1992}),\ \bibinfo {note}
  {unpublished}\BibitemShut {NoStop}%
\bibitem [{\citenamefont {Wita{\l}a}\ and\ \citenamefont
  {Gl{\"o}ckle}(2011)}]{Wit11}%
  \BibitemOpen
  \bibfield  {author} {\bibinfo {author} {\bibfnamefont {H.}~\bibnamefont
  {Wita{\l}a}}\ and\ \bibinfo {author} {\bibfnamefont {W.}~\bibnamefont
  {Gl{\"o}ckle}},\ }\href@noop {} {\bibfield  {journal} {\bibinfo  {journal}
  {Phys. Rev. C}\ }\textbf {\bibinfo {volume} {83}},\ \bibinfo {pages} {034004}
  (\bibinfo {year} {2011})}\BibitemShut {NoStop}%
\bibitem [{\citenamefont {Slaus}\ \emph {et~al.}(1971)\citenamefont {Slaus},
  \citenamefont {Sunier}, \citenamefont {Thompson}, \citenamefont {Young},
  \citenamefont {Verba}, \citenamefont {Margaziotis}, \citenamefont {Doherty},\
  and\ \citenamefont {Cahill}}]{Sla71}%
  \BibitemOpen
  \bibfield  {author} {\bibinfo {author} {\bibfnamefont {I.}~\bibnamefont
  {Slaus}}, \bibinfo {author} {\bibfnamefont {J.~W.}\ \bibnamefont {Sunier}},
  \bibinfo {author} {\bibfnamefont {G.}~\bibnamefont {Thompson}}, \bibinfo
  {author} {\bibfnamefont {J.~C.}\ \bibnamefont {Young}}, \bibinfo {author}
  {\bibfnamefont {J.~W.}\ \bibnamefont {Verba}}, \bibinfo {author}
  {\bibfnamefont {D.~J.}\ \bibnamefont {Margaziotis}}, \bibinfo {author}
  {\bibfnamefont {P.}~\bibnamefont {Doherty}}, \ and\ \bibinfo {author}
  {\bibfnamefont {R.~T.}\ \bibnamefont {Cahill}},\ }\href@noop {} {\bibfield
  {journal} {\bibinfo  {journal} {Phys. Rev. Lett.}\ }\textbf {\bibinfo
  {volume} {26}},\ \bibinfo {pages} {789} (\bibinfo {year} {1971})}\BibitemShut
  {NoStop}%
\bibitem [{\citenamefont {Bovet}\ \emph {et~al.}(1978)\citenamefont {Bovet},
  \citenamefont {Foroughi},\ and\ \citenamefont {Rossel}}]{Bov78}%
  \BibitemOpen
  \bibfield  {author} {\bibinfo {author} {\bibfnamefont {E.}~\bibnamefont
  {Bovet}}, \bibinfo {author} {\bibfnamefont {F.}~\bibnamefont {Foroughi}}, \
  and\ \bibinfo {author} {\bibfnamefont {J.}~\bibnamefont {Rossel}},\
  }\href@noop {} {\bibfield  {journal} {\bibinfo  {journal} {Nucl. Phys. A}\
  }\textbf {\bibinfo {volume} {304}},\ \bibinfo {pages} {29} (\bibinfo {year}
  {1978})}\BibitemShut {NoStop}%
\bibitem [{\citenamefont {Soukup}\ \emph {et~al.}(1979)\citenamefont {Soukup},
  \citenamefont {Cameron}, \citenamefont {Fielding}, \citenamefont {Hussein},
  \citenamefont {Lam},\ and\ \citenamefont {Neilson}}]{Sou79}%
  \BibitemOpen
  \bibfield  {author} {\bibinfo {author} {\bibfnamefont {J.}~\bibnamefont
  {Soukup}}, \bibinfo {author} {\bibfnamefont {J.~M.}\ \bibnamefont {Cameron}},
  \bibinfo {author} {\bibfnamefont {H.~W.}\ \bibnamefont {Fielding}}, \bibinfo
  {author} {\bibfnamefont {A.~H.}\ \bibnamefont {Hussein}}, \bibinfo {author}
  {\bibfnamefont {S.~T.}\ \bibnamefont {Lam}}, \ and\ \bibinfo {author}
  {\bibfnamefont {G.~C.}\ \bibnamefont {Neilson}},\ }\href@noop {} {\bibfield
  {journal} {\bibinfo  {journal} {Nucl. Phys. A}\ }\textbf {\bibinfo {volume}
  {322}},\ \bibinfo {pages} {109} (\bibinfo {year} {1979})}\BibitemShut
  {NoStop}%
\bibitem [{\citenamefont {Guratzsch}\ \emph {et~al.}(1980)\citenamefont
  {Guratzsch}, \citenamefont {K{\"u}hn}, \citenamefont {Kumpf}, \citenamefont
  {M{\"o}sner}, \citenamefont {Neubert}, \citenamefont {Pilz}, \citenamefont
  {Schmidt},\ and\ \citenamefont {Tesch}}]{Gur80}%
  \BibitemOpen
  \bibfield  {author} {\bibinfo {author} {\bibfnamefont {H.}~\bibnamefont
  {Guratzsch}}, \bibinfo {author} {\bibfnamefont {B.}~\bibnamefont {K{\"u}hn}},
  \bibinfo {author} {\bibfnamefont {H.}~\bibnamefont {Kumpf}}, \bibinfo
  {author} {\bibfnamefont {J.}~\bibnamefont {M{\"o}sner}}, \bibinfo {author}
  {\bibfnamefont {W.}~\bibnamefont {Neubert}}, \bibinfo {author} {\bibfnamefont
  {W.}~\bibnamefont {Pilz}}, \bibinfo {author} {\bibfnamefont {G.}~\bibnamefont
  {Schmidt}}, \ and\ \bibinfo {author} {\bibfnamefont {S.}~\bibnamefont
  {Tesch}},\ }\href@noop {} {\bibfield  {journal} {\bibinfo  {journal} {Nucl.
  Phys. A}\ }\textbf {\bibinfo {volume} {342}},\ \bibinfo {pages} {239}
  (\bibinfo {year} {1980})}\BibitemShut {NoStop}%
\bibitem [{\citenamefont {von Witsch}\ \emph {et~al.}(1980)\citenamefont {von
  Witsch}, \citenamefont {\surname{G{\'o}mez} Moreno}, \citenamefont
  {Rosenstock}, \citenamefont {Franke},\ and\ \citenamefont
  {Steinheuer}}]{Wit80}%
  \BibitemOpen
  \bibfield  {author} {\bibinfo {author} {\bibfnamefont {W.}~\bibnamefont {von
  Witsch}}, \bibinfo {author} {\bibfnamefont {B.}~\bibnamefont
  {\surname{G{\'o}mez} Moreno}}, \bibinfo {author} {\bibfnamefont
  {W.}~\bibnamefont {Rosenstock}}, \bibinfo {author} {\bibfnamefont
  {R.}~\bibnamefont {Franke}}, \ and\ \bibinfo {author} {\bibfnamefont
  {B.}~\bibnamefont {Steinheuer}},\ }\href@noop {} {\bibfield  {journal}
  {\bibinfo  {journal} {Phys. Lett.}\ }\textbf {\bibinfo {volume} {91B}},\
  \bibinfo {pages} {342} (\bibinfo {year} {1980})}\BibitemShut {NoStop}%
\bibitem [{\citenamefont {Glasgow}\ \emph {et~al.}(1974)\citenamefont
  {Glasgow}, \citenamefont {Velkley}, \citenamefont {Brandenberger},
  \citenamefont {McEllistrem}, \citenamefont {Hennecke},\ and\ \citenamefont
  {Breitenbecher}}]{Gla74}%
  \BibitemOpen
  \bibfield  {author} {\bibinfo {author} {\bibfnamefont {D.~W.}\ \bibnamefont
  {Glasgow}}, \bibinfo {author} {\bibfnamefont {D.~E.}\ \bibnamefont
  {Velkley}}, \bibinfo {author} {\bibfnamefont {J.~D.}\ \bibnamefont
  {Brandenberger}}, \bibinfo {author} {\bibfnamefont {M.~T.}\ \bibnamefont
  {McEllistrem}}, \bibinfo {author} {\bibfnamefont {H.~J.}\ \bibnamefont
  {Hennecke}}, \ and\ \bibinfo {author} {\bibfnamefont {D.~V.}\ \bibnamefont
  {Breitenbecher}},\ }\href@noop {} {\bibfield  {journal} {\bibinfo  {journal}
  {Nucl. Instr. Methods}\ }\textbf {\bibinfo {volume} {114}},\ \bibinfo {pages}
  {521} (\bibinfo {year} {1974})}\BibitemShut {NoStop}%
\bibitem [{\citenamefont {Finckh}\ \emph {et~al.}(1987)\citenamefont {Finckh},
  \citenamefont {Geissdörfer}, \citenamefont {Lin}, \citenamefont
  {Schindler},\ and\ \citenamefont {Strate}}]{Fin87}%
  \BibitemOpen
  \bibfield  {author} {\bibinfo {author} {\bibfnamefont {E.}~\bibnamefont
  {Finckh}}, \bibinfo {author} {\bibfnamefont {K.}~\bibnamefont
  {Geissdörfer}}, \bibinfo {author} {\bibfnamefont {R.}~\bibnamefont {Lin}},
  \bibinfo {author} {\bibfnamefont {S.}~\bibnamefont {Schindler}}, \ and\
  \bibinfo {author} {\bibfnamefont {J.}~\bibnamefont {Strate}},\ }\href
  {\doibase https://doi.org/10.1016/0168-9002(87)90886-2} {\bibfield  {journal}
  {\bibinfo  {journal} {Nucl. Instr. Methods Phys. Res. A}\ }\textbf {\bibinfo
  {volume} {262}},\ \bibinfo {pages} {441 } (\bibinfo {year}
  {1987})}\BibitemShut {NoStop}%
\bibitem [{\citenamefont {Drosg}\ and\ \citenamefont {Otuka}(2015)}]{Dro15}%
  \BibitemOpen
  \bibfield  {author} {\bibinfo {author} {\bibfnamefont {M.}~\bibnamefont
  {Drosg}}\ and\ \bibinfo {author} {\bibfnamefont {N.}~\bibnamefont {Otuka}},\
  }\href@noop {} {}\bibinfo {type} {Tech. Rep.}\ \bibinfo {number}
  {INDC(AUS)-0019}\ (\bibinfo  {institution} {IAEA Nuclear Data Section},\
  \bibinfo {address} {Vienna, Austria},\ \bibinfo {year} {2015})\BibitemShut
  {NoStop}%
\bibitem [{\citenamefont {Dietze}\ and\ \citenamefont {Klein}(1982)}]{Die82}%
  \BibitemOpen
  \bibfield  {author} {\bibinfo {author} {\bibfnamefont {G.}~\bibnamefont
  {Dietze}}\ and\ \bibinfo {author} {\bibfnamefont {H.}~\bibnamefont {Klein}},\
  }\href@noop {} {}\bibinfo {type} {Tech. Rep.}\ \bibinfo {number} {PTB-ND-22}\
  (\bibinfo  {institution} {Physikalisch-Technische Bundesanstalt},\ \bibinfo
  {address} {Braunschweig, Germany},\ \bibinfo {year} {1982})\BibitemShut
  {NoStop}%
\bibitem [{\citenamefont {Faddeev}(1961)}]{Fad61}%
  \BibitemOpen
  \bibfield  {author} {\bibinfo {author} {\bibfnamefont {L.~D.}\ \bibnamefont
  {Faddeev}},\ }\href@noop {} {\bibfield  {journal} {\bibinfo  {journal}
  {Soviet Physics JETP}\ }\textbf {\bibinfo {volume} {12}},\ \bibinfo {pages}
  {1014} (\bibinfo {year} {1961})}\BibitemShut {NoStop}%
\bibitem [{\citenamefont {Machleidt}\ \emph {et~al.}(1996)\citenamefont
  {Machleidt}, \citenamefont {Sammarruca},\ and\ \citenamefont {Song}}]{Mac96}%
  \BibitemOpen
  \bibfield  {author} {\bibinfo {author} {\bibfnamefont {R.}~\bibnamefont
  {Machleidt}}, \bibinfo {author} {\bibfnamefont {F.}~\bibnamefont
  {Sammarruca}}, \ and\ \bibinfo {author} {\bibfnamefont {Y.}~\bibnamefont
  {Song}},\ }\href@noop {} {\bibfield  {journal} {\bibinfo  {journal} {Phys.
  Rev. C}\ }\textbf {\bibinfo {volume} {53}},\ \bibinfo {pages} {R1483}
  (\bibinfo {year} {1996})}\BibitemShut {NoStop}%
\bibitem [{\citenamefont {Chadwick}\ \emph {et~al.}(2011)\citenamefont
  {Chadwick}, \citenamefont {Herman}, \citenamefont {Oblo{\v z}insk{\' y}}
  \emph {et~al.}}]{Cha11}%
  \BibitemOpen
  \bibfield  {author} {\bibinfo {author} {\bibfnamefont {M.}~\bibnamefont
  {Chadwick}}, \bibinfo {author} {\bibfnamefont {M.}~\bibnamefont {Herman}},
  \bibinfo {author} {\bibfnamefont {P.}~\bibnamefont {Oblo{\v z}insk{\' y}}},
  \emph {et~al.},\ }\href {\doibase 10.1016/j.nds.2011.11.002} {\bibfield
  {journal} {\bibinfo  {journal} {Nucl. Data Sheets}\ }\textbf {\bibinfo
  {volume} {112}},\ \bibinfo {pages} {2887 } (\bibinfo {year} {2011})},\
  \bibinfo {note} {special Issue on ENDF/B-VII.1 Library}\BibitemShut {NoStop}%
\bibitem [{{SAID Database}()}]{SAID}%
  \BibitemOpen
  {SAID Database},\ \href@noop {} {\enquote {\bibinfo {title} {{SAID} partial
  wave analysis},}\ }\bibinfo {howpublished}
  {\url{http://gwdac.phys.gwu.edu/}},\ \bibinfo {note} {accessed
  2017-03-07}\BibitemShut {NoStop}%
\bibitem [{\citenamefont {Glasgow}\ \emph {et~al.}(1976)\citenamefont
  {Glasgow}, \citenamefont {Purser}, \citenamefont {Hogue}, \citenamefont
  {Clement}, \citenamefont {Stelzer}, \citenamefont {Mack}, \citenamefont
  {Boyce}, \citenamefont {Epperson}, \citenamefont {Buccino}, \citenamefont
  {Lisowski}, \citenamefont {Glendinning}, \citenamefont {Bilpuch},
  \citenamefont {Newson},\ and\ \citenamefont {Gould}}]{Gla76}%
  \BibitemOpen
  \bibfield  {author} {\bibinfo {author} {\bibfnamefont {D.~W.}\ \bibnamefont
  {Glasgow}}, \bibinfo {author} {\bibfnamefont {F.~O.}\ \bibnamefont {Purser}},
  \bibinfo {author} {\bibfnamefont {H.}~\bibnamefont {Hogue}}, \bibinfo
  {author} {\bibfnamefont {J.~C.}\ \bibnamefont {Clement}}, \bibinfo {author}
  {\bibfnamefont {K.}~\bibnamefont {Stelzer}}, \bibinfo {author} {\bibfnamefont
  {G.}~\bibnamefont {Mack}}, \bibinfo {author} {\bibfnamefont {J.~R.}\
  \bibnamefont {Boyce}}, \bibinfo {author} {\bibfnamefont {D.~H.}\ \bibnamefont
  {Epperson}}, \bibinfo {author} {\bibfnamefont {S.~G.}\ \bibnamefont
  {Buccino}}, \bibinfo {author} {\bibfnamefont {P.~W.}\ \bibnamefont
  {Lisowski}}, \bibinfo {author} {\bibfnamefont {S.~G.}\ \bibnamefont
  {Glendinning}}, \bibinfo {author} {\bibfnamefont {E.~G.}\ \bibnamefont
  {Bilpuch}}, \bibinfo {author} {\bibfnamefont {H.~W.}\ \bibnamefont {Newson}},
  \ and\ \bibinfo {author} {\bibfnamefont {C.~R.}\ \bibnamefont {Gould}},\
  }\href@noop {} {\bibfield  {journal} {\bibinfo  {journal} {Nucl. Sci. and
  Eng.}\ }\textbf {\bibinfo {volume} {61}},\ \bibinfo {pages} {521} (\bibinfo
  {year} {1976})}\BibitemShut {NoStop}%
\bibitem [{\citenamefont {Drosg}\ and\ \citenamefont {Schwerer}(1987)}]{Dro87}%
  \BibitemOpen
  \bibfield  {author} {\bibinfo {author} {\bibfnamefont {M.}~\bibnamefont
  {Drosg}}\ and\ \bibinfo {author} {\bibfnamefont {O.}~\bibnamefont
  {Schwerer}},\ }in\ \href@noop {} {\emph {\bibinfo {booktitle} {Handbook on
  Nuclear Acitivation Data}}},\ \bibinfo {series and number} {\bibinfo {series}
  {IAEA Tech. Rep. Ser.}\ No.\ \bibinfo {number} {273}},\ \bibinfo {editor}
  {edited by\ \bibinfo {editor} {\bibfnamefont {K.}~\bibnamefont {Okamoto}}}\
  (\bibinfo {address} {Vienna, Austria},\ \bibinfo {year} {1987})\BibitemShut
  {NoStop}%
\bibitem [{\citenamefont {Skibi\ifmmode~\acute{n}\else \'{n}\fi{}ski}\ \emph
  {et~al.}(2018)\citenamefont {Skibi\ifmmode~\acute{n}\else \'{n}\fi{}ski},
  \citenamefont {Volkotrub}, \citenamefont {Golak}, \citenamefont
  {Topolnicki},\ and\ \citenamefont {Wita\l{}a}}]{Ski18}%
  \BibitemOpen
  \bibfield  {author} {\bibinfo {author} {\bibfnamefont {R.}~\bibnamefont
  {Skibi\ifmmode~\acute{n}\else \'{n}\fi{}ski}}, \bibinfo {author}
  {\bibfnamefont {Y.}~\bibnamefont {Volkotrub}}, \bibinfo {author}
  {\bibfnamefont {J.}~\bibnamefont {Golak}}, \bibinfo {author} {\bibfnamefont
  {K.}~\bibnamefont {Topolnicki}}, \ and\ \bibinfo {author} {\bibfnamefont
  {H.}~\bibnamefont {Wita\l{}a}},\ }\href {\doibase 10.1103/PhysRevC.98.014001}
  {\bibfield  {journal} {\bibinfo  {journal} {Phys. Rev. C}\ }\textbf {\bibinfo
  {volume} {98}},\ \bibinfo {pages} {014001} (\bibinfo {year}
  {2018})}\BibitemShut {NoStop}%
\bibitem [{\citenamefont {Ziegler}\ \emph {et~al.}(2010)\citenamefont
  {Ziegler}, \citenamefont {Ziegler},\ and\ \citenamefont {Biersack}}]{SRIM}%
  \BibitemOpen
  \bibfield  {author} {\bibinfo {author} {\bibfnamefont {J.~F.}\ \bibnamefont
  {Ziegler}}, \bibinfo {author} {\bibfnamefont {M.}~\bibnamefont {Ziegler}}, \
  and\ \bibinfo {author} {\bibfnamefont {J.}~\bibnamefont {Biersack}},\ }\href
  {\doibase https://doi.org/10.1016/j.nimb.2010.02.091} {\bibfield  {journal}
  {\bibinfo  {journal} {Nucl. Instr. Methods Phys. Res. B}\ }\textbf {\bibinfo
  {volume} {268}},\ \bibinfo {pages} {1818 } (\bibinfo {year} {2010})},\
  \bibinfo {note} {19th International Conference on Ion Beam
  Analysis}\BibitemShut {NoStop}%
\end{thebibliography}
%

\end{document}